%% file: techreport.tex
\documentclass[english]{article}
\usepackage{lmodern}

\usepackage[T1]{fontenc}
\usepackage[latin9]{inputenc}
\usepackage[a4paper]{geometry}
\usepackage{color}
\usepackage{babel}
\usepackage{amsmath}
\usepackage{amsthm}
\usepackage{amssymb}
\usepackage{stmaryrd}
\usepackage{graphicx}
\usepackage{setspace}
\usepackage{esint}
\usepackage{caption}
\usepackage{subcaption}
\usepackage{natbib}
\usepackage[unicode=true,bookmarks=true,bookmarksnumbered=false,bookmarksopen=false,breaklinks=false,pdfborder={0 0 0},pdfborderstyle={},backref=page,colorlinks=true]{hyperref}
\hypersetup{pdftitle={Models made of modules},pdfauthor={Jacob Murray Holmes Robert},linkcolor=RoyalBlue,citecolor=RoyalBlue}

\makeatletter

 \theoremstyle{definition}
  \newtheorem{example}{\protect\examplename}

\usepackage{babel}
\usepackage{fullpage}

\usepackage[dvipsnames,svgnames,x11names,hyperref]{xcolor}

\newcommand{\argmax}{\operatornamewithlimits{argmax}}

\makeatother

\providecommand{\examplename}{Example}

\begin{document}

\title{Better together? Statistical learning in models made of modules}

\author{Pierre E. Jacob\thanks{Department of Statistics, Harvard University, USA, pjacob@g.harvard.edu}, Lawrence M. Murray\thanks{Department of Information Technology, Uppsala University, Sweden}, 
Chris C. Holmes\thanks{Department of Statistics, and Nuffield Department of Medicine, University of Oxford, UK}, Christian P.  Robert \thanks{CEREMADE, Universit\'e Paris-Dauphine, PSL Research University, France, and Department of Statistics, University of Warwick, UK, xian@ceremade.dauphine.fr}.}

\maketitle

\begin{abstract}
In modern applications, statisticians are faced with integrating heterogeneous data modalities relevant 
for an inference, prediction, or decision problem. In such circumstances, it is convenient 
to use a graphical model to represent the statistical dependencies, via a set of connected `modules', each relating to a specific data modality, and drawing on specific domain expertise in their development. 
In principle, given data, the conventional statistical update then allows for coherent uncertainty
quantification and information propagation through and across the modules. However, misspecification
of any module can contaminate the estimate and update of others, often in unpredictable ways. In various settings, particularly when 
certain modules are trusted more than others, 
practitioners have preferred to avoid learning with the full model in favor of approaches that restrict the information propagation between modules,  
for example by restricting propagation to only particular directions along the edges of the graph. In this article, we investigate
why these modular approaches might be preferable to the full model in 
misspecified settings.  We propose principled criteria to choose between modular
and full-model approaches.  The question arises in many applied settings,
including large stochastic dynamical systems, 
meta-analysis, epidemiological models, air pollution models, pharmacokinetics-pharmacodynamics, and causal inference with propensity scores.
\end{abstract}


\section{Introduction\label{sec:Introduction}}

\subsection{The setting \label{sec:question}}

Consider the situation where a statistical model has been assembled from different components, which we call \emph{modules}. Each of
these may have been developed by a separate community, or built using specific domain knowledge of a particular data modality. Such joint models,
sometimes termed \emph{hierarchical}~\citep{robert2007bayesian}, \emph{super}~\citep{Shen:climatesupermodels}, or \emph{coupled}~\citep{Beal2010} models, are
becoming widespread as measurement technologies and data storage become cheap,
and as efforts to quantify uncertainty intensify. For example, in a model relating air pollution to human health, the joint
model might be made of an air pollution component, guided by climate science
and data from monitoring stations, and a component for human health, based on
medical science and electronic health records \citep[see e.g.][]{blangiardo2011bayesian}. 

In principle, conventional statistical updating
tackles all modules jointly with the advantage that all
uncertainties can be treated simultaneously and coherently. This is
achieved by the posterior distribution in ideal settings \citep{bernardo2009bayesian,gelman2014bayesian}.
However, in a joint model where
information flows both ways between any pair of modules, misspecification of either leads to
misspecification of the full model \citep{liu2009}, potentially leading to
misleading quantification of uncertainties \citep{grunwald2012safe,kleijn2012bernstein,muller2013risk}.
This motivates approaches that depart
from learning with the full model. There may also be other motivations to eschew the full model, such as computational constraints
and data confidentiality.

\begin{figure}[h]
\begin{centering}
    \includegraphics[width=0.5\textwidth]{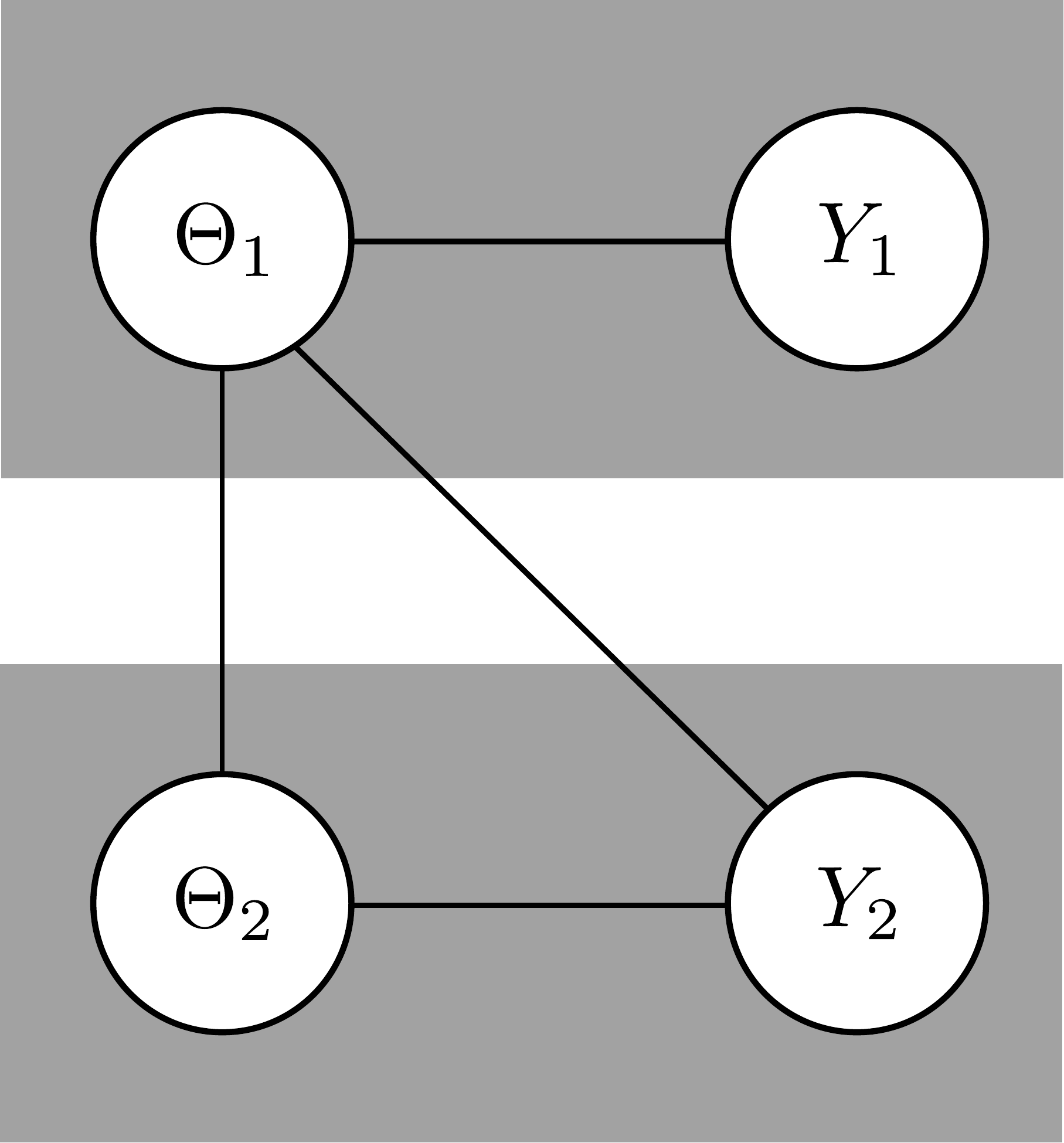}
\par\end{centering}
\caption{\label{fig:Variables}Variables involved in a model made of two modules.
The first module has parameter $\theta_{1}$ and data $Y_{1}$, while
the second is defined conditionally upon $\theta_{1}$, with parameter
$\theta_{2}$ and data $Y_{2}$.}
\end{figure}

To understand the problem and the statistical issues that arise in its simplest
form, consider a graphical model made of just two modules as shown in Figure
\ref{fig:Variables}. In the first module we observe data modality $Y_1$ with a
corresponding likelihood $p_{1}(Y_{1}|\theta_{1})$ parameterized by $\theta_1$.
We will utilize a Bayesian formulation and a prior distribution
$p_{1}(\theta_{1})$. In the absence of other information the inference on
$\theta_{1}$ obtains the posterior distribution
$\pi_{1}(\theta_{1}|Y_{1})\propto p_{1}(Y_{1}|\theta_{1})p_{1}(\theta_{1})$.
Note that, in general, $\theta_1$ is simply an unknown of interest, for example a
realization of a future observable, such that $\pi_{1}(\theta_{1}|Y_{1})$
represents a predictive distribution.  We are interested in the situation where
$\theta_{1}$ is then used in a \emph{second module}, introducing extra
parameters $\theta_{2}$ and data $Y_{2}$. To make the second module
operational, some knowledge on $\theta_{1}$ is required, so that its likelihood
and prior distribution may depend on $\theta_{1}$. The likelihood of this
second module is $p_{2}\left(Y_{2}|\theta_{1},\theta_{2}\right)$, and its prior
distribution $p_{2}(\theta_{2}|\theta_{1})$. When all of the components are
well specified then the joint model provides optimal learning about all of the
unknowns \citep{zellner1988optimal}. However, for a number of reasons---model misspecification,
numerous missing values in certain modalities, contamination of errors,
{\em{a priori}} trust in the specification of some modules more than in others,
computational constraints and data privacy\textemdash one might want to depart
from this full model update.

Departing from the full model then raises some crucial questions such as: 
can we cut the undesired feedback of some components on others
without hampering uncertainty propagation? Can we design principled methods
to decide whether to use the full model or modular approaches? Can we formalise the problem within a valid
Bayesian framework? It is the aim of this article to facilitate answers to such questions and propose a principled way 
to proceed through the use of decision theoretic arguments. Following others \citep[e.g.][]{liu2009}, we refer to the general area of inference in models made of modules as ``modularization''. 

\subsection{Background literature}

Notions of modularization crop up in many applied settings, reviewed below, but the systematic 
statistical evaluation of the techniques has received relatively little attention in 
the methodology literature. Some general issues are described
in \citet{liu2009}, with applications to
computer model calibration. Computational challenges associated
with certain modular approaches are discussed in \citet{plummer2014cuts}.
Both of these articles present reproducible examples which we investigate in Section \ref{sec:Numerical-experiments}. 
In fact the concept of cutting feedback is already implemented in conventional Bayesian software such as WinBUGS which includes 
a `cut function' for multiple imputation and
plug-in or two-step approaches  \citep{liu2009,plummer2014cuts}.
Specific examples of modularization appear in diverse applications, 
such as air pollution \citep{blangiardo2011bayesian},
epidemiological models \citep{maucort2008international,finucane2013estimating,Li2017},
pharmacokinetics-pharmacodynamics \citep{bennett2001errors,lunn2009combining}, 
meta-analysis \citep{lunn:2013,kaizar2015incorporating} and propensity scores \citep{mccandless2010cutting,Zigler:Watts:2013,zigler2014uncertainty}.

The example of pharmacokinetics-pharmacodynamics (PKPD) 
is representative of the strong connection between modularization and misspecification.
Pharmacokinetics (PK) is the study of the body's effect on a drug,
while pharmacodynamics (PD) is the study of a drug's effect on the
body. It is generally believed that the PK part is more precise, or
at least better understood scientifically, than the PD part. 
This motivates, e.g. in \citet{bennett2001errors}, a first module fit
separately on the PK data, and a second module that uses the first
inference in an errors-in-variables model for the PD part. In WinBUGS,
the ``cut'' function is intended for such situations \citep{plummer2014cuts}.
For instance, \citet{lunn2009combining} consider ``cutting''
the feedback of information from variables in the PD module to variables
in the PK module: ``the four models considered {[}corresponding to
various cuts{]} can be thought of as representing varying degrees
of confidence in the PK model relative to the PD model'' (p. 32).
Thus, in PKPD studies, modular approaches are motivated by
the suspected misspecification of the PD module.

Modular approaches are routinely used in econometrics
\citep[e.g.][]{pagan1984econometric,newey1994large,murphy2002estimation}. For
instance, a regression model might be calibrated first. Then the residuals or
the fitted values might be used as covariates in a second regression model.
This is sometimes referred to as generated regressors
\citep{pagan1984econometric} or two-step estimation
\citep{newey1994large,murphy2002estimation}. The latter mentions computational
reasons to motivate a two-step approach, and also notes: ``the researcher may
be reluctant to hypothesize a specific joint distribution for the random
components of the unobservables in the first- and second-step models.'' (p.
88-89).

In climate modeling, modules developed by often-separate scientific communities
are coupled to model the whole Earth system~\citep{Goosse2015}. These include
atmospheric, ocean, land, ice and biogeochemical models. In one such example,
often called \emph{coupled physical-biological models}, physical models of the
ocean are used to force marine biogeochemical models. This is usually achieved
by taking a single representative trajectory from the physical model and
plugging it into the biological model, but there is increasing interest in
considering how uncertainty in the physics may propagate through to uncertainty
in the biology, and how informative the biological observations may be on the
physics~\cite[see e.g.][]{Cossarini2009,Beal2010,Mattern2013}.

In \citet{woodard2013hierarchical}, the authors describe a regression
using a nonparametric representation of functional predictors. Independently
for a number of individuals $i\in\left\{ 1,\ldots,n\right\} $, a
function $f_{i}$ is estimated, which involves a number of subject-specific
parameters $\theta_{i}$. The parameters $\theta_{i}$ are then used
in a regression of an outcome $Y_{i}$ across individuals (Equation
(5) in \citet{woodard2013hierarchical}, which then takes the form
of a spline regression). In this article, there is an interest in
taking the uncertainty of $\theta$ into account when performing the
regression, but in such a way that ``any potential misspecification of the regression
model (5) does not negatively affect estimation of the subject functions
$f_{i}$'' (p. 11).

When considering a Bayesian modelling of $k$-nearest neighbour classification,
\citet{titterington2006knn} cut the missing class labels at the predictive
sites from the genuine parameters of their model in order to avoid being
swamped by the imprecision on these labels. The parameters are first estimated
based on the observed class labels and the prediction is then operated
conditional on this first step.

In the context of air pollution, \citet{blangiardo2011bayesian} propose an empirical comparison
between a fully Bayesian approach and a modularized solution.  
A modularized approach is also presented in \citet{finucane2013estimating},
for the estimation of the prevalence of transmitted HIV drug resistance.
In modeling linkage disequilibrium among multiple SNPs, \citet{li2003modeling}
describe a modularized approach, ``although {[}the full Bayesian
approach{]} would be our preferred approach''. There, the modularized
approach seems preferred for computational reasons and the authors randomize over the modular architecture. 
The notion of feedback cutting also appears in the comments of \citet{rougier2008comment} on \citet{sanso2008inferring},
in the context of climate systems, as well as in \citet{sham2012inferring}.

\subsection{Outline and objectives}

Our primary goal is to open a discourse on the statistical issues surrounding
modularization in modern applications, in part by tying together its use across
diverse problem domains. We consider criteria for deciding whether or not to
update using a conventional full modelling approach.  The proposed criteria use
the available training data to quantify the relative merits of the joint and
modular approaches.  Decision theory is a principled framework to address these
issues, and the logarithmic scoring rule provides a default utility function
for modularization, with strong connections to model selection criteria and
Bayes factors as used in conventional Bayesian statistics. 

In Section \ref{sec:frameworkmethods} we consider, in depth, the modularization issues 
arising in the simplified model structure illustrated in Figure \ref{fig:Variables}. 
Since the full model approach is often considered the gold standard in Bayesian
statistics, we discuss in detail in Section \ref{sec:Why-not-fullposterior}
specific reasons why modular approaches might perform better, in the context of model
misspecification. Section \ref{sec:Numerical-experiments} presents
four reproducible examples where modular approaches outperform the full model including the case of 
meta-analysis (Section \ref{sec:metaanalysis}) which goes beyond the setting of a model with two modules. 
Section \ref{sec:Computational-challenges} discusses 
the computational challenges that arise from modularized inference, which may further motivate one particular learning approach
over another. Section \ref{sec:Discussion} provides a short 
conclusion.

\section{Choosing between full models and modularized approaches\label{sec:frameworkmethods}}

In this section, we introduce the notation for a model made of two modules (Section \ref{sec:notation}),
and describe various approaches to statistical inference beyond the full model learning (Section \ref{sub:candidate-modules}).
After having introduced basic elements of decision theory, we describe our proposed criterion to decide whether 
to use the full model or modular approaches in Section \ref{sub:decision:Comparing-distributions-using-decision}. We
summarize our proposed plan of action in Section \ref{sub:predictionplan}.

\subsection{Model with two modules \label{sec:notation}}

The concept of combining multiple sources of information in order to improve decision making or estimation is central 
to statistics. In the context of a model made of modules it helps to distil the problem down to just two modules with two 
sources of information and a common parameter set, as illustrated in Figure \ref{fig:Variables}. Some of the resulting canonical inference problems are given below.
\begin{itemize}
    \item[-] The module of interest might be $(\theta_{1},Y_{1})$. The data $Y_{2}$
        represent some extra data made available to update the inference on
        $\theta_{1}$, through a model that involves a parameter $\theta_{2}$
        that can be considered a nuisance parameter. This will be the setting
        of the example of Section \ref{sub:Biased-data}.
    \item[-] The module of interest might be $(\theta_{2},Y_{2})$, but the model
        involves an unknown parameter $\theta_{1}$, to be learned with data
        $Y_{1}$, which then can be considered a nuisance parameter. This is for instance the case where the second model is 
        a regression of some outcome on covariates, some of which themselves
        predicted from a first model. An example is provided by causal inference with propensity
        scores, as in Section \ref{sub:Propensity-score}. Note that, due to the dependence on $\theta_1$, this case is not symmetric to the previous case.
    \item[-] The first module $(\theta_{1},Y_{1})$ might be of interest for a
        certain community, and the second module $(\theta_{2},Y_{2})$ for another community. Examples arise in, for example, coupled physical-biological ocean models, where $(\theta_{1},Y_{1})$ is a physical model for the dynamics of temperature and salinity $\theta_{1}$ based on data $Y_{1}$, and $(\theta_{2},Y_{2})$ is a biological model for the dynamics of plankton populations $\theta_{2}$ based on data $Y_{2}$~\citep[see e.g.][]{Cossarini2009,Beal2010, Mattern2013}. In this
        case, $\theta_{1}$ is critical in the inference on $\theta_{2}$, along with propagation of uncertainty from $\theta_{1}$ to $\theta_{2}$, but it might be expected that $Y_{2}$ brings little extra information on $\theta_{1}$ given $Y_{1}$. Another example is provided in Section \ref{sub:Epidemiological-study}, where the first model estimates human papillomavirus prevalence, while the second relates this prevalence to cervical cancer incidence \citep{maucort2008international}.

\end{itemize}
The above specification of likelihoods and prior distributions uniquely
defines a joint distribution on $\theta_{1}$, $\theta_{2}$, $Y_{1}$
and $Y_{2}$. We denote the parameter by $(\theta_{1},\theta_{2})$
with prior $p_{1}(\theta_{1})p_{2}(\theta_{2}|\theta_{1})$, the data
by $(Y_{1},Y_{2})$ and the likelihood by $(\theta_{1},\theta_{2})\mapsto p_{1}(Y_{1}|\theta_{1})p_{2}(Y_{2}|\theta_{1},\theta_{2})$.
We refer to this model as the \emph{full model} and the posterior
distribution $\bar{\pi}(\theta_{1},\theta_{2}|Y_{1},Y_{2})$ as the
\emph{full posterior}, with density 
\begin{align}
\bar{\pi}\left(\theta_{1},\theta_{2}|Y_{1},Y_{2}\right) & =\bar{\pi}\left(Y_{1},Y_{2}\right)^{-1}\,p_{2}(Y_{2}|\theta_{1},\theta_{2})p_{2}\left(\theta_{2}|\theta_{1}\right)p_{1}\left(\theta_{1}\right)p_{1}(Y_{1}|\theta_{1})\nonumber \\
 & =\bar{\pi}\left(\theta_{2}|\theta_{1},Y_{2}\right)\bar{\pi}\left(\theta_{1}|Y_{1},Y_{2}\right).\label{eq:fullposterior:joint}
\end{align}
We denote by $n_{1}$ (respectively $n_{2}$) the number of observations
in $Y_{1}$ (respectively $Y_{2}$). The dimension of $\theta_{1}$
(respectively $\theta_{2}$) is denoted $d_{1}$ (respectively $d_{2}$).
We denote by $\bar{\pi}$ any expectation with respect to the full
model, for instance 
\[
\bar{\pi}(Y_{1})=\int p_{1}(Y_{1}|\theta_{1})p_{1}(\theta_{1})d\theta_{1},\quad\text{or}\quad\bar{\pi}(Y_{1},Y_{2})=\int p_{2}(Y_{2}|\theta_{1},\theta_{2})p_{2}\left(\theta_{2}|\theta_{1}\right)p_{1}\left(\theta_{1}\right)p_{1}(Y_{1}|\theta_{1})d\theta_{1}d\theta_{2}.
\]
We also write $\bar{\pi}(\theta_{1}|Y_{1})$ instead of $\pi_{1}(\theta_{1}|Y_{1})$
for the posterior in the first module, $\bar{\pi}(\theta_{1},\theta_{2})$
for the joint prior, etc. For a number of reasons, as stated in Section \ref{sec:question}, one might want to depart from this full
model. 

\subsection{Candidate distributions \label{sub:candidate-modules}}

The full model is one possible assembly of the two modules into a
coherent model. Alternatives exist. We refer to any distribution representing
beliefs on $\theta_{1}$ (or $\theta_{2}$, or both) as a \emph{candidate
distribution} for $\theta_{1}$ (or $\theta_{2}$, or both). We first
enumerate a number of such candidates. These are derived from conceptual
or pragmatic considerations, where in many cases one module is of
primary concern, while the other is only of secondary importance.

\subsubsection{The first module}

For a focus on the first module, such as inferring $\theta_{1}$ or
predicting $Y_{1}$, the full model provides the marginal distribution
$\bar{\pi}(\theta_{1}|Y_{1},Y_{2})$, with density 
\begin{align}
\underbrace{\bar{\pi}(\theta_{1}|Y_{1},Y_{2})}_{\text{full posterior}}%
 & =\frac{\bar{\pi}(Y_{1})}{\bar{\pi}(Y_{1},Y_{2})}\,\underbrace{\bar{\pi}(\theta_{1}|Y_{1})}_{\text{first posterior}}\,\underbrace{\bar{\pi}\left(Y_{2}|\theta_{1}\right)}_{\text{feedback}}.\label{eq:fullposterior:1}
\end{align}
The \emph{feedback} term is 
\begin{equation}
\bar{\pi}\left(Y_{2}|\theta_{1}\right)=\int p_{2}(Y_{2}|\theta_{1},\theta_{2})p_{2}\left(\theta_{2}|\theta_{1}\right)d\theta_{2}.\label{eq:feedbackterm}
\end{equation}
An alternative assembly is to ignore the second module altogether,
and use the inference obtained from the first posterior, $\bar{\pi}(\theta_{1}|Y_{1})$,
only. Starting from Eq. \eqref{eq:fullposterior:1}, this amounts
to neglecting the feedback term, a decision sometimes referred to
as \emph{cutting feedback} \citep{liu2009}. Since $\bar{\pi}(\theta_{1}|Y_{1},Y_{2})$
uses $Y_{2}$ while $\bar{\pi}(\theta_{1}|Y_{1})$ does not, it might
seem intuitively preferable to use the full posterior $\bar{\pi}(\theta_{1}|Y_{1},Y_{2})$.
We will see that this is not necessarily the case, in Sections \ref{sec:Why-not-fullposterior}-\ref{sec:Numerical-experiments}.
Therefore, $\bar{\pi}(\theta_{1}|Y_{1})$ is one possible alternative
to $\bar{\pi}(\theta_{1}|Y_{1},Y_{2})$. Furthermore, the prior distribution
$\bar{\pi}(\theta_{1})$, ignoring both $Y_{1}$ and $Y_{2}$, is
also a candidate to be considered, and so is the posterior distribution of $\theta_{1}$
given $Y_{2}$, ignoring $Y_{1}$.

\subsubsection{The second module}

If the focus is on the second module, the full model provides the
distribution $\bar{\pi}(\theta_{1},\theta_{2}|Y_{1},Y_{2})$ as in
Eq.~\eqref{eq:fullposterior:joint}, with marginal $\bar{\pi}(\theta_{2}|Y_{1},Y_{2})$,
and this is the obvious first candidate. Multiple alternatives to
infer either $\theta_{2}$, or $(\theta_{1},\theta_{2})$, are possible.

Perhaps the simplest is the two-step approach. In the first step,
$\theta_{1}$ is estimated from $Y_{1}$ using $\bar{\pi}(\theta_{1}|Y_{1})$,
and summarized by a point estimate $\hat{\theta}_{1}$. In the second
step, $\hat{\theta}_{1}$ is plugged into the second module, leading
to the distribution 
\begin{equation}
\bar{\pi}(\theta_{2}|\hat{\theta}_{1},Y_{2})=\frac{p_{2}(Y_{2}|\hat{\theta}_{1},\theta_{2})p_{2}(\theta_{2}|\hat{\theta}_{1})}{\bar{\pi}(Y_{2}|\hat{\theta}_{1})}.\label{eq:twostage:ptestimate}
\end{equation}
Equivalently, we can replace $\bar{\pi}(\theta_{1}|Y_{1},Y_{2})$
by $\delta_{\hat{\theta}_{1}}(\theta_{1})$ in Eq. (\ref{eq:fullposterior:joint}),
leading to a joint distribution with density $(\theta_{1},\theta_{2})\mapsto\delta_{\hat{\theta}_{1}}(\theta_{1})\bar{\pi}(\theta_{2}|\hat{\theta}_{1},Y_{2})$,
denoted by $\bar{\pi}(\theta_{2}|\hat{\theta}_{1},Y_{2})$. The uncertainty
on $\theta_{1}$ from the first module is not propagated to the estimation
of $\theta_{2}$, and the second data $Y_{2}$ is not used in the
inference on $\theta_{1}$.

We might want to propagate the uncertainty of the first module, without
accepting feedback of $Y_{2}$ on $\theta_{1}$. This is achieved
by an approach implemented in OpenBUGS and JAGS as the \emph{cut}
function \citep{lunn2000winbugs,plummer2014cuts}. It consists in
replacing $\bar{\pi}(\theta_{1}|Y_{1},Y_{2})$ with $\bar{\pi}(\theta_{1}|Y_{1})$
in Eq. (\ref{eq:fullposterior:joint}), yielding the cut distribution
with density 
\begin{equation}
\pi^{\text{cut}}\left(\theta_{1},\theta_{2}|Y_{1},Y_{2}\right)=\bar{\pi}\left(\theta_{2}|\theta_{1},Y_{2}\right)\bar{\pi}(\theta_{1}|Y_{1})=\frac{\bar{\pi}(Y_{1},Y_{2})}{\bar{\pi}(Y_{1})}\frac{\bar{\pi}\left(\theta_{1},\theta_{2}|Y_{1},Y_{2}\right)}{\bar{\pi}\left(Y_{2}|\theta_{1}\right)}.\label{eq:jointcut}
\end{equation}
The cut distribution is a valid probability distribution that takes
the uncertainty about $\theta_{1}$ into account, while cutting the
feedback of $Y_{2}$ on $\theta_{1}$, in the sense that the marginal
posterior distribution of $\theta_{1}$ is still $\bar{\pi}(\theta_{1}|Y_{1})$.
It can be seen as a probabilistic version of a two-step estimator \citep{newey1994large}.
Candidates on the second module are thus: the full posterior,
the prior, the two-step approach, the cut approach, and the posterior distribution
of $(\theta_{1},\theta_{2})$ given $Y_{2}$ but not $Y_{1}$.

\subsection{Decision-theoretical view\label{sub:decision:Comparing-distributions-using-decision}}

Having introduced candidate distributions for $\theta_{1}$ and for
$(\theta_{1},\theta_{2})$, we now turn to the main question: how
do we choose the most appropriate candidate? The main reason not to
automatically use the full posterior distribution is \emph{model misspecification}.
There are other reasons\textemdash related to computation and privacy,
for example\textemdash but we put these aside for now.

\subsubsection{Optimal actions}

Our approach is to adopt a decision theoretic argument similar to that used for 
Bayesian model comparison in the
misspecified setting (also known as the \emph{M-open} setting) as
described e.g. in \citet{bernardo2009bayesian}.
In a generic parameter inference setting, to avoid confusion with
the notation introduced in the previous section, denote a model by
$M$, a prior by $\text{prior}(\theta|M)$, a likelihood by $\text{likelihood}(Y|\theta,M)$
for some data $Y$ and the posterior by $\text{posterior}(\theta|Y,M)$.
Denote by $p_{\star}$ the data-generating distribution of $Y$. A
model is \emph{misspecified} if there is no $\theta^{\star}$ such
that $\text{likelihood}(Y|\theta^{\star},M)$ is the same distribution as $p_{\star}$.
We introduce a utility function $(\omega,d)\mapsto u(\omega,d)$, where
$\omega$ refers to some unknown state of interest, and $d$ to a
decision or action, such as providing a prediction or choosing to select one of the models. 
Under a data-generating distribution $p_{\star}(\omega)$ on the
unknown states, the expected utility of $d$ is given by $\int u(\omega,d)p_{\star}(\omega)d\omega$.
We introduce a further distribution $p(\omega|\theta,M)$ relating the unknown
states $\omega$ to parameters. The distribution of $\omega$ given
the model and the data $Y$ has density $p(\omega|Y,M)=\int p(\omega|\theta,M)\text{posterior}(\theta|Y,M)d\theta$.
In the misspecified setting, computing the optimal action $d_{\star}$
maximizing $\int u(\omega,d)p_{\star}(\omega)d\omega$ is not feasible.
A practical approach consists in considering the $M$-optimal action
$d_{M}$, defined as 
\[
d_{M}=\argmax_{d}\int u(\omega,d)p(\omega|Y,M)d\omega,
\]
where $p_{\star}$ is replaced by $p(\omega|Y,M)$, leading to a potentially
tractable optimization program. To compare different models, we can
compare the performances of the associated $M$-optimal actions, given
by their expected utility under $p_{\star}$, $u_{M}=\int u(\omega,d_{M})p_{\star}(\omega)d\omega$.
Still, this integral is intractable, but we can envision various approximations
based on data. A standard approach is to split the available data
into a training set, on which the optimal actions $d_{M}$ are computed,
and a test set, used to approximate $u_{M}$ by an empirical average.
Related procedures include the sequential predictive approach described
in Section \ref{sub:prequential}.

In the context of modularization, instead of models, multiple candidate
distributions are available from different modular architectures. We compare them following the same rationale.
Dropping the model index $M$ from the notation, introduce a link
function $p(\omega|\theta)$ relating the parameter $\theta$ to the
unknown state of interest $\omega$. For a candidate $\pi(\theta)$,
the $\pi$-optimal action is 
\[
d_{\pi}=\argmax_{d}\int\int u(\omega,d)p(\omega|\theta)\pi(\theta)d\theta d\omega,
\]
and the associated expected utility is $u_{\pi}=\int u(\omega,d_{\pi})p_{\star}(\omega)d\omega$.
We can then compare the expected utility of different candidates.

The choice of utility functions is potentially arduous and we discuss
the use of predictive criteria as a default choice in Section \ref{sub:predictive-as-default}.
Although we do not see the decision-theoretic framework as controversial
in itself---it underpins most statistical methods---it has direct, and possibly surprising, consequences. For instance,
we will see in Section \ref{sec:Why-not-fullposterior} that the prior
distribution might prove to be better than the posterior distribution in terms
of expected utility, when the task is probabilistic prediction and
the loss function is the logarithmic scoring rule.

\subsubsection{Prediction and logarithmic scoring rule \label{sub:predictive-as-default}}

Ideally an appropriate utility function is available for the problem
at hand. For instance, one would typically know whether their interest
lies in the first or the second module, whether the interest lies
in predicting future observations or not, etc. However, it can be hard to formulate
a utility function that is both faithful to the scientific question
and computationally tractable, and thus, we propose a default choice.
Our choice is related to what the posterior distribution and maximum
likelihood estimator achieve, whether or not a decision-theoretic
framework is explicitly introduced. Recall that the Kullback-Leibler
(KL) divergence from a distribution with density $p^{\prime}$, to
another distribution with density $p$, is denoted by 
\[
\text{KL}(p,p^{\prime})=\int\log\left(\frac{p(y)}{p^{\prime}(y)}\right)p(y)dy.
\]
For a predictive distribution with density $y\mapsto\pi(y)$ and an
observation $Y$, the logarithmic score is $-\log\pi(Y)$. The logarithmic
score satisfies desirable properties to assess predictive distributions
\citep{bernardo2009bayesian}, but other choices are available \citep{gneiting2007strictly,parry2012proper,dawid2015bayesian}.

Using the generic notation of the previous section, $\text{posterior}(\theta|Y)$ is the minimizer of 
\begin{equation}
\int\left(-\log\text{likelihood}(Y|\theta)\right)\nu(\theta)d\theta+\text{KL}\left(\nu(\theta),\text{prior}(\theta)\right),\label{eq:nonasymptoticcriterion}
\end{equation}
over all choices of $\nu(\theta)$ such that the above quantity exists;
see e.g. \citet{bissiri2016general} for more justification of this
optimization program based on coherency arguments.

Indeed, if $\nu$ is the posterior, then Eq. \eqref{eq:nonasymptoticcriterion}
equals $-\log p\left(Y\right)=-\log\int\text{likelihood}(Y|\theta)\text{prior}(\theta)d\theta$.
For any other distribution $\nu$, Eq. \eqref{eq:nonasymptoticcriterion}
gives 
\begin{align*}
\int\left(-\log\text{likelihood}(Y|\theta)\right)\nu(\theta)d\theta+\int\left(\log\frac{\nu(\theta)}{\text{prior}(\theta)}\right)\nu(\theta)d\theta= & -\int\log\frac{\text{prior}(\theta)\text{likelihood}(Y|\theta)}{\nu(\theta)}\nu(\theta)d\theta\\
> & -\log\left(\int\frac{\text{prior}(\theta)\text{likelihood}(Y|\theta)}{\nu(\theta)}\nu(\theta)d\theta\right)\\
 & =-\log p(Y).
\end{align*}
The strict inequality comes from Jensen on strictly concave functions,
the minus sign, and the fact that if $\nu$ is not the posterior,
then $\text{prior}(\theta)\text{likelihood}(Y|\theta)/\nu(\theta)$
is not almost surely constant under $\theta\sim\nu$. Hence any other
$\nu$ yields a larger objective in Eq. \eqref{eq:nonasymptoticcriterion} than the posterior.

From Eq. \eqref{eq:nonasymptoticcriterion}, the posterior puts mass on parameters $\theta$ such that
$-\log\text{likelihood}(Y|\theta)$ is large subject to
KL similarity to the prior, and 
the quantity $-\log\text{likelihood}(Y|\theta)$
has an interpretation as a predictive score. This view on
the posterior holds under misspecification.
Asymptotically in the number of observations, the posterior distribution
(and similarly the maximum likelihood estimator) concentrates around
the parameter value $\theta^{\star}$ that minimizes $\text{KL}(p_{\star},\text{likelihood}(y|\theta))$,
under weak conditions (see, e.g., \citet{kleijn2012bernstein} for Bayesian
asymptotic results). Minimizing that KL is equivalent to minimizing
$\theta\mapsto\int-\log\text{likelihood}(y|\theta)p_{\star}(y)dy$,
the expected loss associated with predicting $Y$ with $y\mapsto\text{likelihood}(y|\theta)$
when $Y\sim p_{\star}$.

Therefore, the task of predicting observations under the logarithmic
score is embedded in likelihood-based approaches and we choose it
as a default. We define the unknown state $\omega$ to be a future observation
$\tilde{y}$, the actions to be probability distributions $q$ on
$\mathbb{Y}$, the utility to be minus the logarithmic scoring rule
$u(\tilde{y},q)=\log q(\tilde{y})$. The link function $p(\tilde{y}|\theta)$
is taken to be the model likelihood. For a candidate distribution
$\pi$ on $\theta$, the $\pi$-optimal action is the predictive distribution
$\pi_{Y}(y)=\int\text{likelihood}(y|\theta)\pi(\theta)d\theta$. Its
expected utility under $p_{\star}$ is given by $\int_{\mathbb{Y}}\log\pi_{Y}(y)p_{\star}(y)dy$,
which is equal to $-\text{KL}(p_{\star},\pi_{Y})$
up to an additive constant.

\subsubsection{Proposed predictive scores\label{sub:prequential}}

Computing the expected utility $u_{\pi}$ associated with a candidate
distribution $\pi$ involves an integral with respect to $p_{\star}$,
which is intractable. Typically, one can come up with the predictive
distribution $\pi_{Y}$ associated with a candidate $\pi$ through
Monte Carlo approximations, but the integral $\int_{\mathbb{Y}}\log\pi_{Y}(y)p_{\star}(dy)$
is out of reach. This type of intractability can be addressed by splitting
the data into training and test sets, by cross-validation, or by a
sequential predictive approach. In this section, we focus on the latter,
also called the prequential approach \citep{dawid1984present}.

The performance of a candidate distribution
can be evaluated sequentially over the data, while updating the distribution
with the same data. We denote by $Y^{1:n}=(Y^{1},\ldots,Y^{n})$ the
$n$ observations which compose the data. In the generic notation
of the previous sections, we introduce a sequence of intermediate
posteriors, denoted $\text{posterior}(\theta|Y^{1:i})$ for $i=1,\ldots,n$,
and associated predictive distributions with density $y\mapsto\int\text{likelihood}(y|\theta)\text{posterior}(\theta|Y^{1:i})d\theta$.
We set $\text{posterior}(\theta|Y^{1:i})$ to be the prior if $i=0$.
Evaluating their predictive performance and summing the scores over
the data yields 
\begin{align}
    \text{[posterior score]}\quad\quad\sum_{i=1}^{n}\log\left(\int\text{likelihood}(Y^{i}|\theta)\text{posterior}(\theta|Y^{1:i-1})d\theta\right) & =\log p(Y^{1:n}),\label{eq:prequential:evidence}
\end{align}
where $p(Y^{1:n})=\int\text{likelihood}(Y^{1:n}|\theta)\text{prior}(\theta)d\theta$.
We retrieve this marginal likelihood, also called the evidence, as
a way of scoring a model \citep{bernardo2009bayesian}; or here, of scoring a prior candidate distribution.
Cutting the feedback of the observations
on the parameter would yield a sequential criterion such as
\begin{equation}
\text{[prior score]}\quad\quad\sum_{i=1}^{n}\log\left(\int\text{likelihood}(Y^{i}|\theta)\text{prior}(\theta)d\theta\right)=\sum_{i=1}^{n}\log p(Y^{i}),\label{eq:prequential:nofeedback}
\end{equation}
which corresponds to repeatedly using the prior prediction for each
new observation, without updating the distribution. Under misspecification,
the score $\sum_{i=1}^{n}\log p(Y^{i})$ might be larger than $\log p(Y^{1:n})$,
even asymptotically in the number of observations; see Section \ref{sec:Why-not-fullposterior}. 
We also introduce a similar predictive criterion for the predictive
performance of the cut distribution. We write the two data sets $Y_{1}=Y_{1}^{1:n_{1}}$
and $Y_{2}=Y_{2}^{1:n_{2}}$. 
The cut score is defined as 
\begin{equation}
\text{[cut score]}\quad\quad    \sum_{i=1}^{n_{2}}\log\left(\int p_{2}(Y_{2}^{i}|\theta_{1},\theta_{2})\pi^\text{cut}(\theta_1,\theta_2|Y_1,Y_2^{1:i-1})d\theta_{1}d\theta_{2}\right),\label{eq:prequential:partialfeedback}
\end{equation}
where $\pi^\text{cut}(\theta_1,\theta_2|Y_1,Y_2^{1:i-1}) = \bar\pi(\theta_1,\theta_2|Y_1)$ if $i=1$. Note that the cut distribution itself
is invariant by re-ordering of $Y_2$ (for i.i.d. data), but the cut score is not.
Therefore, one might prefer to average the cut score over permutations of $Y_2$.

In the first module, prior candidates are $\bar{\pi}(\theta_{1})$
and $\bar{\pi}(\theta_{1}|Y_{2})$. Following the prequential approach
with feedback from $Y_{1}$, the associated predictive scores are
given respectively by $\log\bar{\pi}(Y_{1})$
and $\log\bar{\pi}(Y_{1}|Y_{2})$,
respectively. If we consider only these two scores, we end up comparing
$\bar{\pi}(Y_{1})$ to $\bar{\pi}(Y_{1}|Y_{2})$, which is reminiscent
of a Bayes factor. We can also consider the prior prediction performance
as given by Eq. \eqref{eq:prequential:nofeedback}. Importantly, we should
not compare directly $\bar{\pi}(Y_{1})$ and $\bar{\pi}(Y_{1},Y_{2})$,
as these two quantities correspond to the task of predicting different data sets $Y_1$ and $(Y_1,Y_2)$ and thus
are not commensurate.

Likewise, we can compute scores for the task of predicting $Y_{2}$.
Allowing feedback from $Y_{2}$, we can compare various priors on
$(\theta_{1},\theta_{2})$, such as $\delta_{\hat{\theta}_{1}}\left(\theta_{1}\right)p_{2}(\theta_{2}|\theta_{1})$
and $\bar{\pi}(\theta_{1},\theta_{2}|Y_{1})$, leading respectively
to $\log\bar{\pi}(Y_{2}|\hat{\theta}_{1})$
and $\log\bar{\pi}(Y_{2}|Y_{1})$.
Ignoring $Y_{1}$ but allowing full feedback from $Y_{2}$, the prior
$\bar{\pi}(\theta_{1},\theta_{2})$ would lead to the score $\log\bar{\pi}(Y_{2})$.
We can also envision prior prediction scores without feedback, as
given by Eq. \eqref{eq:prequential:nofeedback}, for any of the prior candidates.
Finally, feedback of $Y_{2}$ on $\theta_{2}$ but not on $\theta_{1}$
leads to the cut score of Eq. \eqref{eq:prequential:partialfeedback},
starting from the prior $\bar{\pi}(\theta_{1},\theta_{2}|Y_{1})$,
and we could also consider similar scores starting from other priors.

\subsection{Plan of action\label{sub:predictionplan}}

If the interest is purely in predictions of $Y_{1}$, $Y_{2}$, or both, then the
plan of action is to compute the predictive scores described above,
and to select candidate distributions corresponding to the highest scores.  The
number of candidates to compare might be daunting, especially if more than two
modules are considered. Practical aspects and intuition on a case-by-case basis
might help reduce the number of scores to compute.

Crucially, if the interest lies in parameter inference, the above plan of action
can lead to problematic decisions. Indeed, the interpretability of parameters might change
when considered as part of a module or as part of another. For instance,
consider the parameter $\theta_{1}$ of the first module. The specification of the likelihood $p_{1}(Y_{1}|\theta_{1})$
assigns some meaning to the parameter, e.g. a location, a scale, or a
regression coefficient. Since $\theta_{1}$ further appears in the
likelihood of the second module, $p_{2}(Y_{2}|\theta_{1},\theta_{2})$,
it is also assigned another interpretation. In the context of model
misspecification, there might be a mismatch between both interpretations.
If we had instead used the notation $\eta_{1}$ for the parameter in the first
module, and $(\theta_{1},\theta_{2})$ for the parameters in the second
module, then the fact that the meaning of $\eta_{1}$ might not generally
coincide with the meaning of $\theta_{1}$ would be more apparent.
In other words, equating the meaning of $\eta_{1}$ to that of $\theta_{1}$
is an extra assumption that, in general, should be challenged; 
related discussions can be found in the concrete examples of Section \ref{sec:Numerical-experiments}.

We propose a plan of action that assumes that the meaning
of $\theta_{1}$ is as intended in the specification of the first
module.
\begin{itemize}
    \item[-] In the first module, for each candidate $\pi(\theta_{1})$, compute
        the corresponding score as described in Section \ref{sub:prequential}.
        Select the candidate distribution on $\theta_{1}$ that yields the
        most accurate predictions of $Y_{1}$ according to the scores, and denote it by $\pi_{1}^{\star}(\theta_{1})$.
    \item[-] Choose among the candidate distributions on $(\theta_{1},\theta_{2})$
        that admit $\pi_{1}^{\star}(\theta_{1})$ as a first marginal distribution,
        by computing the corresponding predictive scores for $Y_{2}$, as described in Section \ref{sub:prequential}.
\end{itemize}
This plan action will be tested on four examples in Section \ref{sec:Numerical-experiments}.

\section{Modular approaches can outperform the full posterior \label{sec:Why-not-fullposterior}}

Since we propose to choose amongst a set of candidates, only
one of which is the conventional posterior distribution in the full model, it is worth reflecting on some of
the reasons why the full model may not be optimal. Here we provide
some discussion and examples, starting with a comparison between the prior and posterior distributions,
in a misspecified setting.

\subsection{Prior versus posterior\label{subsec:Prior-versus-posterior}}

Consider again the generic notation introduced in Section \ref{sub:decision:Comparing-distributions-using-decision}.
The posterior distribution, $\text{posterior}(\theta|y)$, is expected
to concentrate toward $\theta^{\star}$ that minimizes $\mathrm{KL}(p_{\star},\text{likelihood}(y|\theta))$.
In the well-specified case, the minimal KL divergence is zero, so that
the posterior predictive distribution is asymptotically optimal in
terms of expected utility for the logarithmic scoring rule. In the
misspecified case, this is no longer true. For a candidate $\pi(\theta)$,
the expected score is $\int\log\left(\int\text{likelihood}(y|\theta)\pi(\theta)d\theta\right)p_{\star}(y)dy$,
which might be larger than $\int\log\text{likelihood}(y|\theta^{\star})p_{\star}(y)dy$,
the expected score associated with the predictive distribution $\text{likelihood}(y|\theta^{\star})$.
One such candidate may the prior distribution, $\pi(\theta)=\text{prior}(\theta)$.
Intuitively, mixing over various parameters might lead to better predictive
power than conditioning on any single parameter value, even the apparently
optimal parameter value $\theta^{\star}$, in the misspecified case.
\begin{example}
    \label{example:posteriorworsethanprior} Consider a prior with density
    $\text{prior}(\theta)=\varphi(\theta;0,1)$ and a likelihood $\text{likelihood}(y|\theta)=\varphi(y;\theta,1)$,
    where $x\mapsto\varphi(x;\mu,\sigma^{2})$ denotes the pdf of the
    Normal distribution $\mathcal{N}(\mu,\sigma^{2})$. If $p_{\star}$
    is such that $\int|y|p_{\star}(y)dy<\infty$, the posterior concentrates
    around the mean of $p_{\star}$ as $n\to\infty$. Assume that this
    mean is zero. The posterior predictive converges to $\mathcal{N}(0,1)$,
    while the prior prediction is $\mathcal{N}(0,2)$. For various possible
    distributions $p_{\star}$ with zero mean, the prior prediction is
    closer to $p_{\star}$ than the posterior predictive, with respect
    to KL divergence or any sensible metric. In particular, if $p_{\star}$ itself is $\mathcal{N}(0,2)$,
    then the prior prediction cannot be outperformed.
\end{example}

The loss of predictive power can be detected by computing prequential
criteria, as proposed in Section \ref{sub:prequential}. For instance,
if we score the prior predictions with Eq. \eqref{eq:prequential:nofeedback},
and the posterior predictions with Eq. \eqref{eq:prequential:evidence},
then, after normalizing both quantities by $n$, the former goes to
\[\int\log\left(\int\text{likelihood}(y|\theta)\text{prior}(\theta)d\theta\right)p_{\star}(y)dy,\]
while the latter goes to 
\[\int\log\text{likelihood}(y|\theta^{\star})p_{\star}(y)dy.\]
In other words, laws of large numbers can be used to approximate expected
utilities with respect to $p_{\star}$, at least asymptotically in
the number of observations, without having to split the data into
training and test sets. 

Further insight is available from Eq. \eqref{eq:nonasymptoticcriterion}.
The posterior finds parameters $\theta$ such that $\log\text{likelihood}(Y|\theta)$
is large, under prior similarity constraints. However, prediction 
can be done without conditioning on only one parameter; instead we can use a predictive
distribution $\pi_{Y}(y)=\int\text{likelihood}(y|\theta)\pi(\theta)d\theta$
which mixes over $\theta$ according to a candidate $\pi$. The predictive
performance might then be better than for any single choice of $\theta$.
Nothing prevents, in general, the strict inequality in 
\[
    \text{KL}(p_{\star}(y),\int\text{likelihood}(y|\theta)\pi(\theta)d\theta)<\min_{\theta}\text{KL}(p_{\star}(y),\text{likelihood}(y|\theta)),
\]
for some choice of $\pi$, as illustrated in Example \ref{example:posteriorworsethanprior}
above. Note that, in other misspecified cases, some parameter $\theta^{\star}$
might indeed lead to better predictions than any mixing of parameters.
To summarize, in misspecified settings, the posterior distribution
can be better or worse than the prior in terms of predictive performance,
which motivates the development of methods to choose whether to use
the posterior distribution or not.

\subsection{Modular versus full\label{subsec:Modular-versus-full}}

Since the posterior is not in general superior to the prior under
model misspecification, for the task of prediction, it is perhaps
not surprising that the full posterior is not always superior to modular
approaches. By the same argument as in Eq. \eqref{eq:nonasymptoticcriterion},
the full posterior minimizes
\begin{align}
 & \int\left(-\log p_{1}(Y_{1}|\theta_{1})-\log p_{2}(Y_{2}|\theta_{1},\theta_{2})\right)\nu(\theta_{1},\theta_{2})d\theta_{1}d\theta_{2}+\text{KL}\left(\nu,\bar{\pi}(\theta_{1},\theta_{2})\right),\label{eq:optimprogram:fullposterior}
\end{align}
over all distributions $\nu$ on $(\theta_{1},\theta_{2})$. In terms
of predicting $Y_{1}$, it is convenient to write that the full posterior
minimizes$\int\left(-\log p_{1}(Y_{1}|\theta_{1})\right)\nu(\theta_{1})d\theta_{1}+\text{KL}\left(\nu,\bar{\pi}(\theta_{1},\theta_{2}|Y_{2})\right),$
over all distributions $\nu$ on $\theta_{1}$. The issue might be
in the KL similarity term with the distribution $\bar{\pi}(\theta_{1},\theta_{2}|Y_{2})$,
which might not be an appealing prior if the second module is misspecified.
In particular, that prior might contradict the prior $\bar{\pi}(\theta_{1})$
that was originally specified for $\theta_{1}$. Similar reasonings
can be done for the prediction of $Y_{2}$: we might dispute the appeal
of $\bar{\pi}(\theta_{1},\theta_{2}|Y_{1})$ as a prior distribution
on $(\theta_{1},\theta_{2})$, if the first likelihood $p_{1}(Y_{1}|\theta_{1})$
is misspecified, compared to the prior $\bar{\pi}(\theta_{1},\theta_{2})$.

In terms of predicting both $Y_{1}$ and $Y_{2}$, there are other
alternatives to the predictive derived from the full posterior. In
particular, we may wish to weight the predictive scores corresponding
to $Y_{1}$ and $Y_{2}$. If we replace $-\log p_{1}(Y_{1}|\theta_{1})-\log p_{2}(Y_{2}|\theta_{1},\theta_{2})$
in Eq. \eqref{eq:optimprogram:fullposterior} by a weighted sum $-\gamma_{1}\log p_{1}(Y_{1}|\theta_{1})-\gamma_{2}\log p_{2}(Y_{2}|\theta_{1},\theta_{2})$,
the solution of the minimization program has a density proportional
to 
\[
(\theta_{1},\theta_{2})\mapsto p_{1}(\theta_{1})p_{2}(\theta_{2}|\theta_{1})p_{1}(Y_{1}|\theta_{1})^{\gamma_{1}}p_{2}(Y_{2}|\theta_{1},\theta_{2})^{\gamma_{2}}.
\]
Reasons to weight the terms include the fact that the two quantities
$\log p_{1}(Y_{1}|\theta_{1})$ and $\log p_{2}(Y_{2}|\theta_{1},\theta_{2})$
are not necessarily commensurate, being based on different data sets
and/or different models. The choice of weights $(\gamma_{1},\gamma_{2})$
could reflect some suspicion of misspecification of some modules compared
to others. The choice of weights $(\gamma_{1},\gamma_{2})$ is discussed
e.g. in \citet{holmes2017assigning}. Putting the likelihood to some
power, or replacing it by other functions in case of misspecified
models, has been found increasingly useful \citep{zhang2006,grunwald2012safe,muller2013risk,bissiri2016general}.

Finally, we note that the full posterior has an asymptotic advantage
over the plug-in approach in terms of predicting $Y_{2}$. Indeed,
the plug-in distribution $\bar{\pi}(\theta_{2}|\hat{\theta}_{1},Y_{2})$
minimizes
\[
\int\left(-\log p_{2}(Y_{2}|\hat{\theta}_{1},\theta_{2})\right)\nu(\theta_{2})d\theta_{2}+\text{KL}\left(\nu,p_{2}(\theta_{2}|\hat{\theta}_{1})\right),
\]
over all distributions $\nu$ on $\theta_{2}$. Asymptotically in
$n_{2}$, the plug-in distribution might concentrate on some $\hat{\theta}_{2}$
that minimizes $\theta_{2}\mapsto\text{KL}(p_{\star}(y_{2}),p_{2}(y_{2}|\hat{\theta}_{1},\theta_{2}))$.
Then, $(\hat{\theta}_{1},\hat{\theta}_{2})$ will be different than
the pair $(\theta_{1}^{\star},\theta_{2}^{\star})$ that minimizes
$(\theta_{1},\theta_{2})\mapsto\text{KL}(p_{\star}(y_{2}),p_{2}(y_{2}|\theta_{1},\theta_{2}))$,
unless it happens that $\hat{\theta}_{1}$ and $\theta_{1}^{\star}$
coincide. Since the optimization is over a larger set in the latter
case, the predictive performance of $(\theta_{1}^{\star},\theta_{2}^{\star})$
is in general superior to that of $(\hat{\theta}_{1},\hat{\theta}_{2})$.
Therefore, we can expect worse asymptotic predictive performance for
$Y_{2}$ when using the plug-in approach compared to the full posterior.
Since mixing over the parameter $\theta_{1}$ could improve the predictive
perfomance, the cut distribution might lead to better predictions
than the plug-in approach. The cut distribution might perform either
worse or better than the full posterior in terms of predictive performance
of $Y_{2}$, even asymptotically in $n_{2}$. Indeed, the cut distribution
maintains some averaging over the parameters, and thus can possibly
lead to better predictions than the ones obtained by conditioning
on $(\theta_{1}^{\star},\theta_{2}^{\star})$ only, 
as discussed in \ref{subsec:Prior-versus-posterior}.

\section{Numerical experiments\label{sec:Numerical-experiments}}

We consider four examples from the statistics literature \citep[namely][]{liu2009,plummer2014cuts,zigler2016central} where modular
approaches are described and motivated as an alternative to the full posterior. 
We investigate whether our proposed method to choose between modular and full model inference
(as summarized in Section \ref{sub:predictionplan}) confirms or contradicts the literature. 
To the best of our knowledge, our method provides the first quantitative way of guiding this choice.
Computational methods used to produce the tables and figures of this section 
are described in Section \ref{sec:Computational-challenges}.


\subsection{Biased data\label{sub:Biased-data}}

The first example is borrowed from \citet{liu2009}, 
where the emphasis is on the existence of situations where the full posterior 
behaves in undesirable ways compared to modular approaches.
We use the example to check whether our proposed method automatically selects modular approaches.

Assume that the data $Y_{1}=(Y_{1}^{1},\ldots,Y_{1}^{n_{1}})$
are independent Normal variables, $Y_{1}^{i}\sim\mathcal{N}(\theta_{1},1)$
for all $1\leq i\leq n_{1}$. A prior distribution $\mathcal{N}(0,\lambda_{1}^{-1})$
is specified on $\theta_{1}$, where $\lambda_{1}$ denotes precision.
This defines the first module. We are given extra data, denoted $Y_{2}=(Y_{2}^{1},\ldots,Y_{2}^{n_{2}})$,
perhaps in large quantity but suspected to be biased. We assume $Y_{2}^{i}\sim\mathcal{N}(\theta_{1}+\theta_{2},1)$
for all $1\leq i\leq n_{2}$, where $\theta_{2}$ is the unknown bias.
The prior distribution on $\theta_{2}$ is $\mathcal{N}(0,\lambda_{2}^{-1})$,
which concludes the specification of the second module. We generate
data with $\theta_{1}^{\star}=0$, $n_{1}=100$, $\theta_{2}^{\star}=1$
and $n_{2}=1000$, reflecting a large bias in the extra data. Furthermore,
we use $\lambda_{1}=1$ and $\lambda_{2}=100$ (i.e. a standard deviation
of $0.1$), reflecting over-optimism in the size of the bias. For
our particular realization of the dataset, we observe a mean of $Y_{1}$
approximately equal to $0.06$ and a mean of $Y_{2}$ approximately
equal to $1.04$.

As described in \citet{liu2009}, parameter estimation is not necessarily
improved by using the full model over modular approaches. We consider
predictive criteria to decide whether to use the full model or not.
Table \ref{tab:biaseddata} contains predictive scores for 
$Y_{1}$ and $Y_{2}$, under various candidates. 

\begin{figure}
\begin{centering}
    \includegraphics[width=0.5\textwidth]{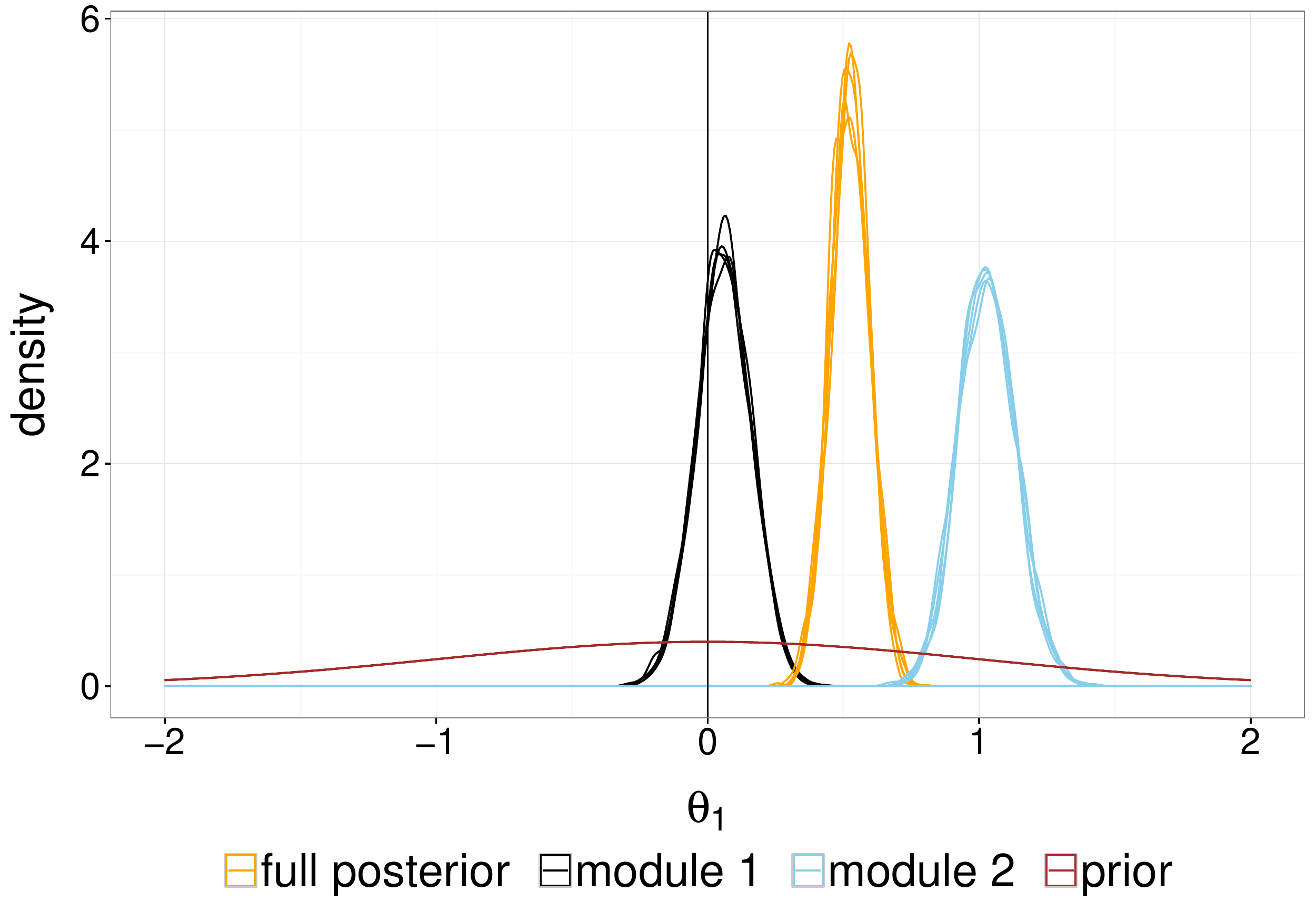}\includegraphics[width=0.5\textwidth]{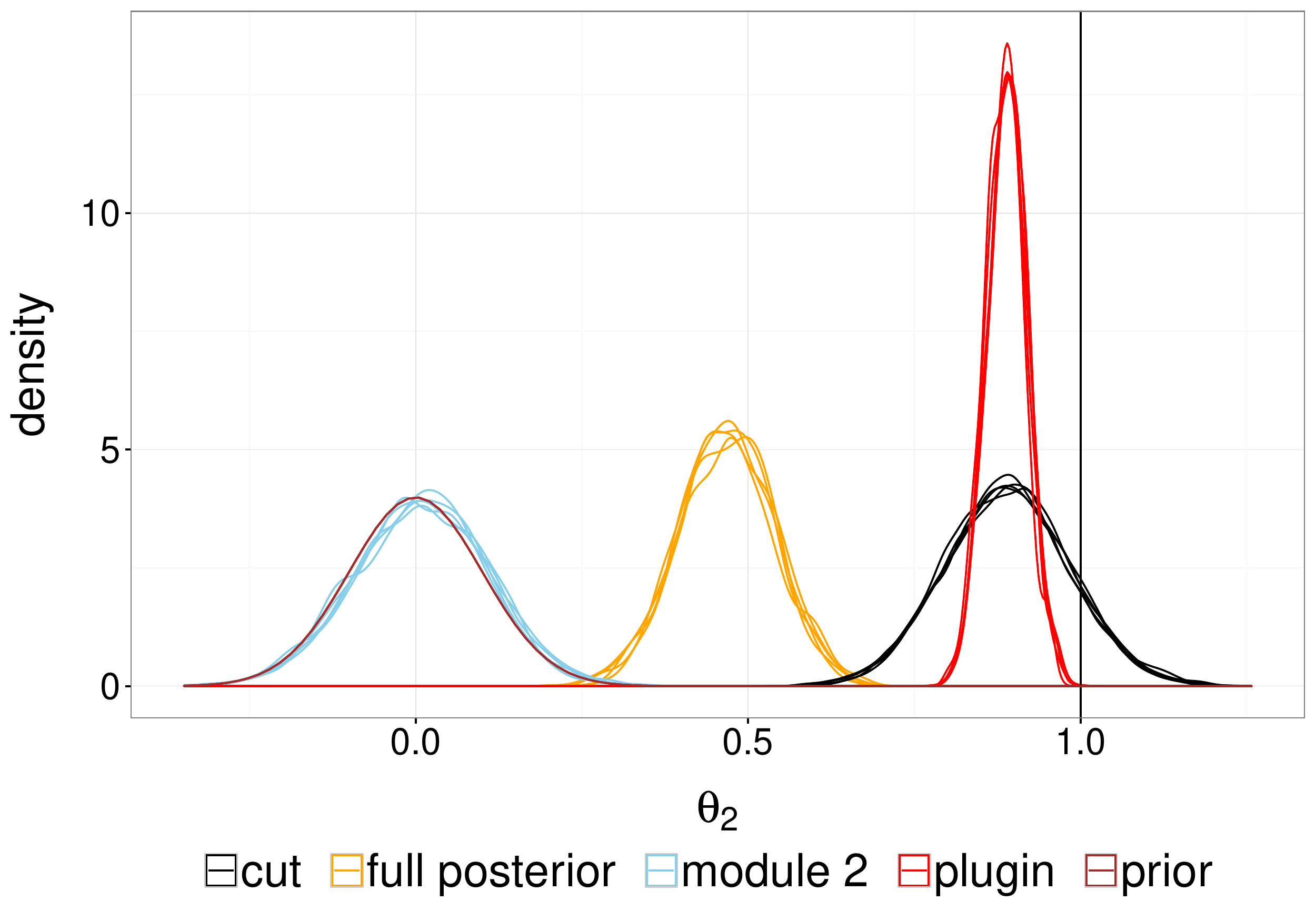}
\par\end{centering}
\caption{\label{fig:biaseddata:marginalparameters}Various candidate distributions
for $\theta_{1}$ (left) and $\theta_{2}$ (right), in the biased
data example of Section \ref{sub:Biased-data}. The data-generating
values are indicated by vertical black lines.}
\end{figure}

For the task of predicting $Y_{1}$, 
the full model has worse predictive performance than the first module
on its own. In fact, the prior distribution has better predictive
power than the full posterior. The posterior in the second module
only, $\bar{\pi}(\theta_{1}|Y_{2})$, leads to the worse predictive
performance. The marginal distributions are shown in Figure \ref{fig:biaseddata:marginalparameters}
(left). We can see from the plot why $\bar{\pi}(\theta_{1}|Y_{1})$
is more satisfactory than the other candidate distributions. In terms
of interpretation, the first module specifies $\theta_{1}$ as the
location of $Y_{1}$, whereas the second module specifies $\theta_{1}+\theta_{2}$
as the location of $Y_{2}$. This is different from the intended interpretation,
which is that $\theta_{1}$ remains the location of $Y_{1}$, while $\theta_{2}$ quantifies bias, that is, the location of $Y_{2}-Y_{1}$.

For the task of predicting $Y_{2}$, 
the full model has worse predictive performance than the second module
on its own, but better performance than the cut approach, which itself
performs similarly to the plug-in approach, where $\theta_{1}$ is
replaced by the expectation of $\bar{\pi}(\theta_{1}|Y_{1})$. Thus,
to predict $Y_{2}$, the best option is to ignore $Y_{1}$ and to
use the candidate $\pi(\theta_{1},\theta_{2}|Y_{2})$. However, to
interpret the parameters, we would follow the plan of action of Section
\ref{sub:predictionplan}, and use the cut distribution which
has $\bar{\pi}(\theta_{1}|Y_{1})$ as its first marginal, because
that distribution is best at predicting $Y_{1}$. Note that
we would choose the cut distribution without looking at 
the predictive scores for $Y_2$, since it is the only
candidate with $\bar{\pi}(\theta_{1}|Y_{1})$ as its first marginal. In general,
there could be multiple candidates with $\bar{\pi}(\theta_{1}|Y_{1})$ as a first marginal.

\begin{figure}
\begin{centering}
    \includegraphics[width=0.5\textwidth]{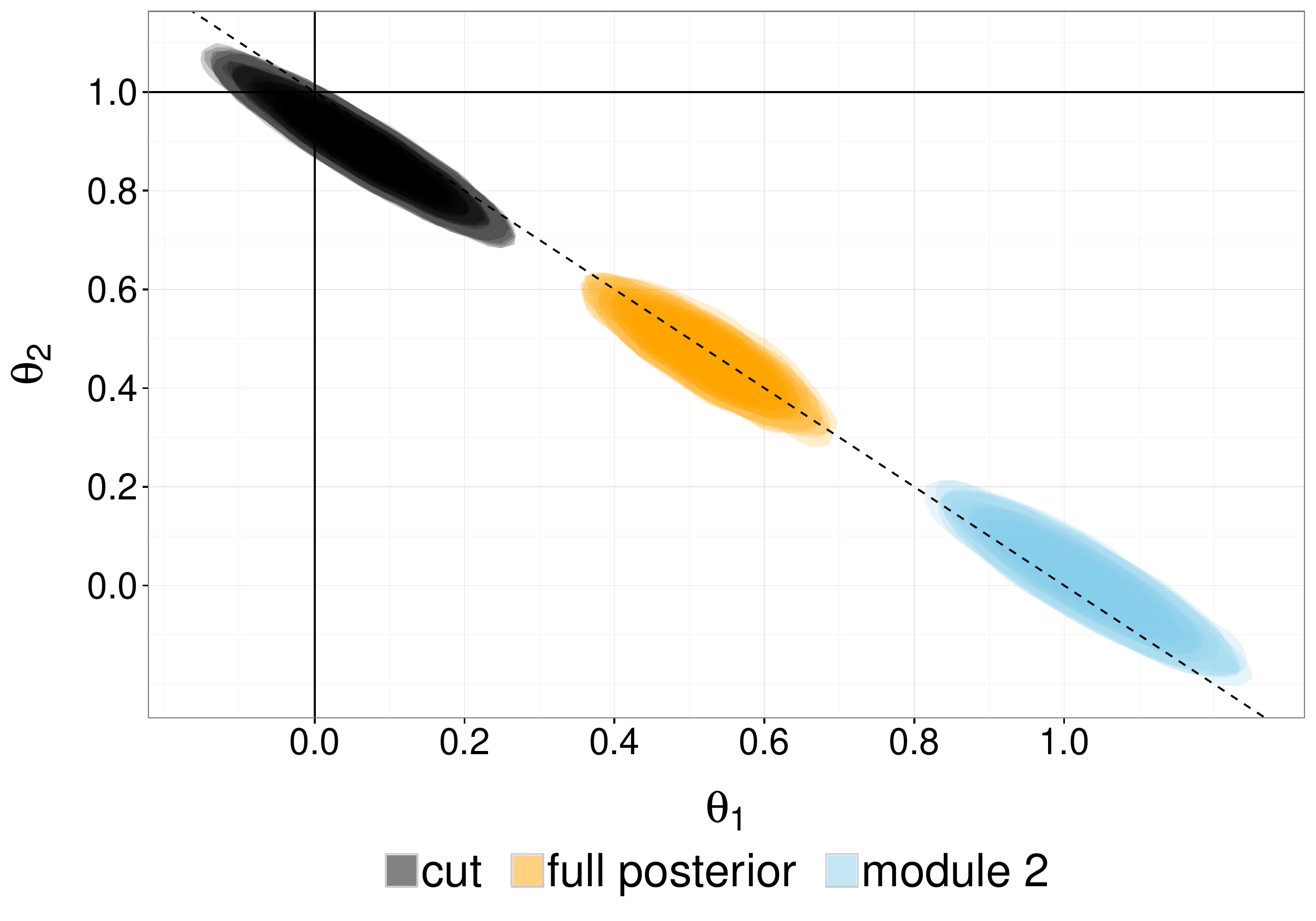}\includegraphics[width=0.5\textwidth]{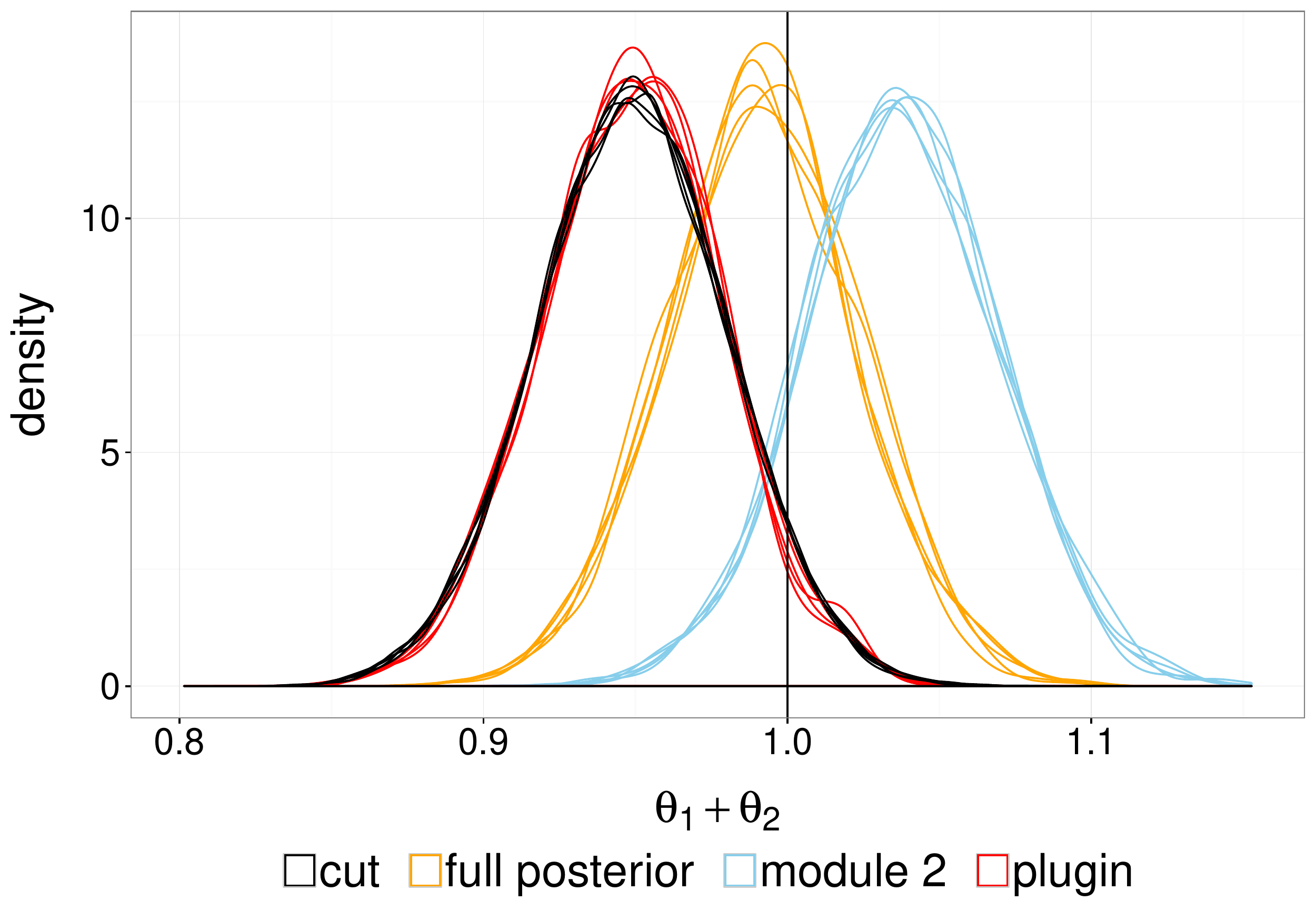}
\par\end{centering}
\caption{\label{fig:biasedata:bivariate}Cut distribution, full posterior and
posterior in the second module, in the biased data example of Section
\ref{sub:Biased-data}. Joint distribution on the left, and marginal
distribution of $\theta_{1}+\theta_{2}$ on the right. The data-generating
values are indicated by black lines. On the left, the diagonal dashed
line indicates the set of parameters such that $\theta_{1}+\theta_{2}=1$,
which are optimal for prediction.}
\end{figure}

The marginal distributions
of $\theta_{2}$ are shown in Figure \ref{fig:biaseddata:marginalparameters}
(right). Here, we can check that the cut distribution seems the most
satisfactory in terms of parameter inference, since we know the data-generating
values.
The joint distributions of $(\theta_{1},\theta_{2})$ under the cut,
the full posterior and the posterior under module 2 only are shown
on the left in Figure \ref{fig:biasedata:bivariate}. The plug-in
approach is excluded as it only provides degenerate joint distributions.
We see that all three distributions have concentrated around the set
$\left\{ (\theta_{1},\theta_{2}):\quad\theta_{1}+\theta_{2}=1\right\} $,
since the data-generating distribution  of $Y_2$ is $\mathcal{N}(1,1)$.
Only the cut distribution puts most of its mass around the values
$(\theta_{1}^{\star},\theta_{2}^{\star})$. The right-most plot in
Figure \ref{fig:biasedata:bivariate} shows the marginal distributions
of $\theta_{1}+\theta_{2}$; we see that the marginal resulting from
the full posterior is most concentrated around $1$, which is the
optimal value for predicting $Y_{2}$.

\begin{table}
\begin{centering}
\begin{minipage}[t]{0.4\columnwidth}%
\input{biaseddata.predictY1}\end{minipage}\hspace{1cm}%
\begin{minipage}[t]{0.4\columnwidth}%
\input{biaseddata.predictY2}\end{minipage}
\par\end{centering}
\caption{\label{tab:biaseddata}Predictive performance of various candidates
in the biased data example of Section \ref{sub:Biased-data}, in terms
of predicting $Y_{1}$ (left) and predicting $Y_{2}$ (right). The
numbers represent the average score over $5$ Monte Carlo runs, with
minimum and maximum values in brackets.}

\end{table}

\subsection{Epidemiological study\label{sub:Epidemiological-study}}

\citet{plummer2014cuts} describes a simple and reproducible example
inspired by epidemiological studies, and in particular by an investigation
of the international correlation between human papillomavirus (HPV)
prevalence and cervical cancer incidence \citep{maucort2008international}.
The focus of \citet{plummer2014cuts} is on computational challenges
with the cut distribution (see Section \ref{sec:Computational-challenges}),
while we use the example to test whether our proposed method selects the full
posterior or the cut distribution.

In the toy version of the model, the first module specifies HPV prevalence,
independently for datasets collected in $13$ countries. The parameter
$\theta_{1}=\theta_{1,1:13}$ has prior distribution $\theta_{1,i}\sim\text{Beta}(1,1)$,
independently for all $1\leq i\leq13$. The data $Y_{1}^{1:13}$ are
$13$ pairs of integers, the first being the number of women infected
with high-risk HPV, and the second being a population size; we write
$Y_{1}^{i}=(Y_{1}^{i}[1],Y_{1}^{i}[2])$ for all $1\leq i\leq13$.
The likelihood specifies that the data are independent across countries
and that $Y_{1}^{i}[1]\sim\text{Binomial}(Y_{1}^{i}[2],\theta_{1,i})$
for all $1\leq i\leq13$. By conjugacy, the posterior distribution
factorizes into a product of Beta distributions, $\text{Beta}(1+Y_{1}^{i}[1],1+Y_{1}^{i}[2]-Y_{1}^{i}[1])$
for all $1\leq i\leq13$.

The second module represents the relationship between HPV prevalence
$\theta_{1}$ and cancer incidence, in the form of a Poisson regression.
The parameters are $\theta_{2}=(\theta_{2,1},\theta_{2,2})$, both
assumed to follow a Normal distribution $\mathcal{N}(0,10^{3})$,
a priori. The second module specifies 
\[
\forall i\in\left\{ 1,\ldots,13\right\} \quad Y_{2}^{i}[1]\sim\text{Poisson}(\exp\left(\theta_{2,1}+\theta_{1,i}\theta_{2,2}+Y_{2}^{i}[2]\right)),
\]
where the second dataset $Y_{2}^{1:13}$ is made of pairs $\left(Y_{2}^{i}[1],Y_{2}^{i}[2]\right)_{i=1}^{13}$
representing, respectively, the number of cancer incidents and the number
of years at follow-up. It is suspected that the Poisson regression
might be misspecified, and \citet{plummer2014cuts} discusses computational
methods to approximate the cut distribution. Note that the first parameters
are the covariates in the regression specified by the second module.
Therefore, the parameters have no clear interpretation if the second
module is considered on its own: one would not typically consider
both covariates and regression coefficients to be unknown simultaneously.

\begin{figure}
\begin{centering}
    \includegraphics[width=0.5\textwidth]{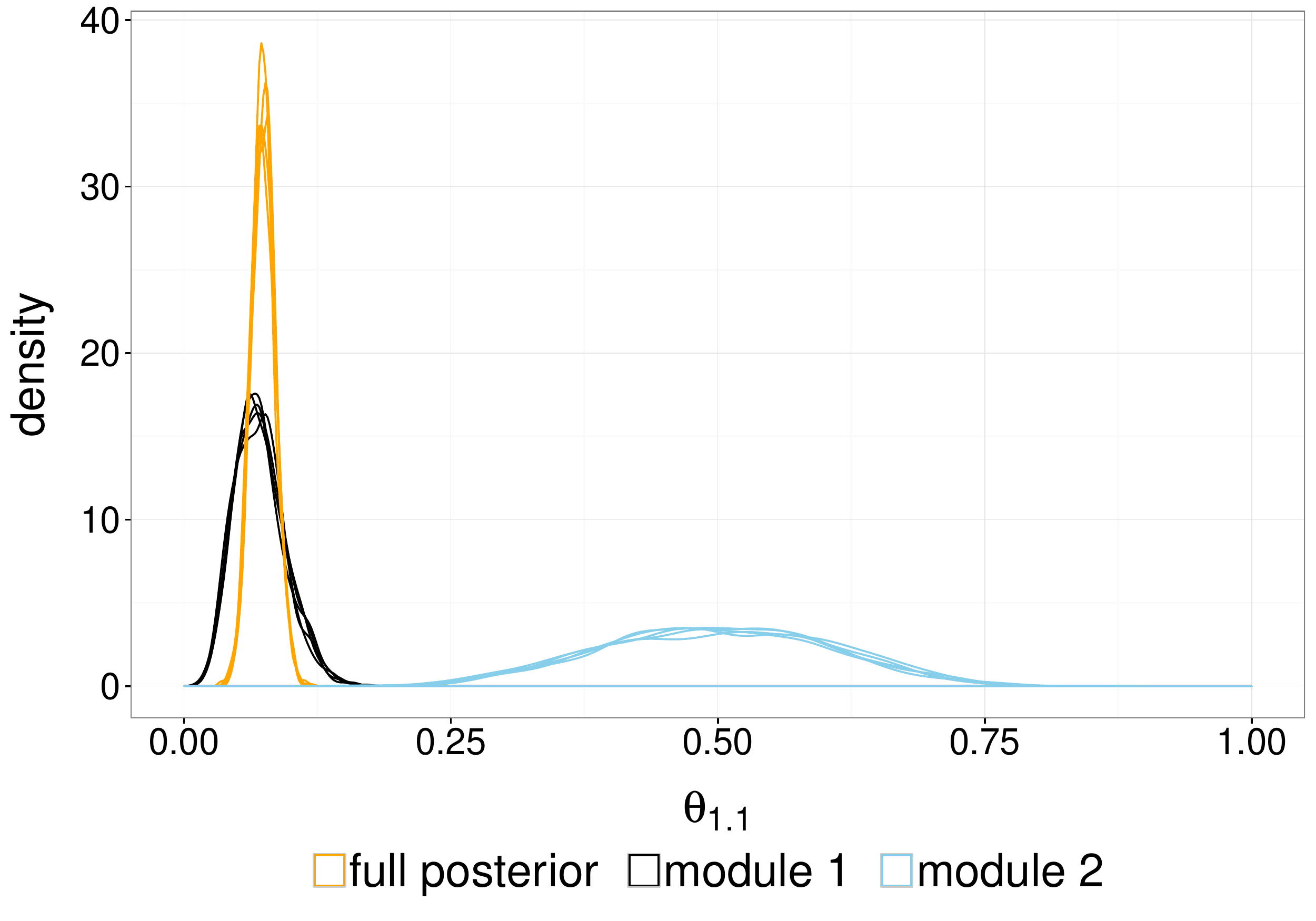}\includegraphics[width=0.5\textwidth]{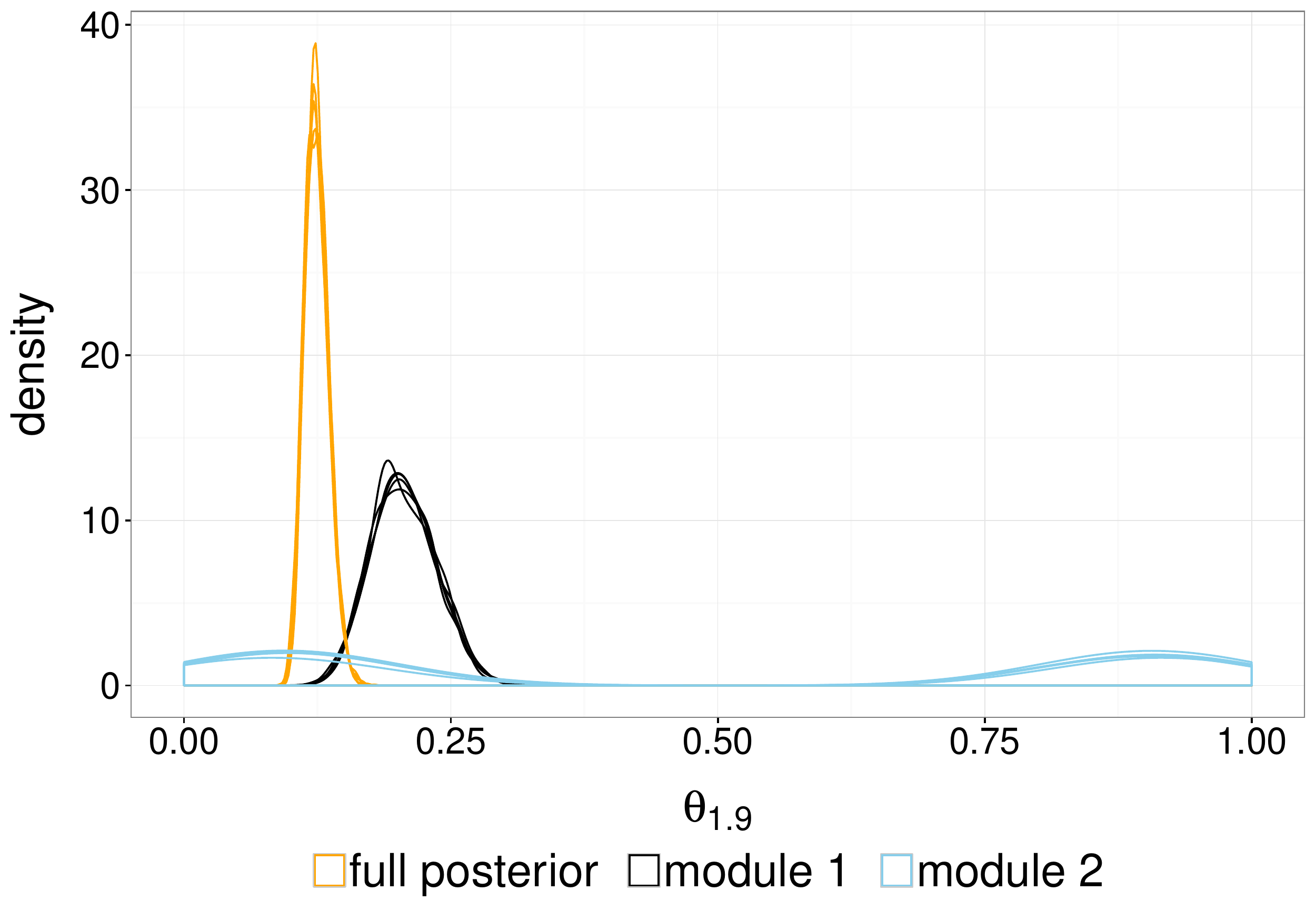}
\par\end{centering}
\caption{\label{fig:plummer:marginalparameters1}Various candidate distributions
for $\theta_{1,1}$ (left) and $\theta_{1,9}$ (right), in the epidemiological
study of Section \ref{sub:Epidemiological-study}. The full posterior
is in agreement with the first module's posterior for some parameters
(such as $\theta_{1,1}$) but not for others (such as $\theta_{1,9}$).}
\end{figure}

\begin{figure}
\begin{centering}
    \includegraphics[width=0.5\textwidth]{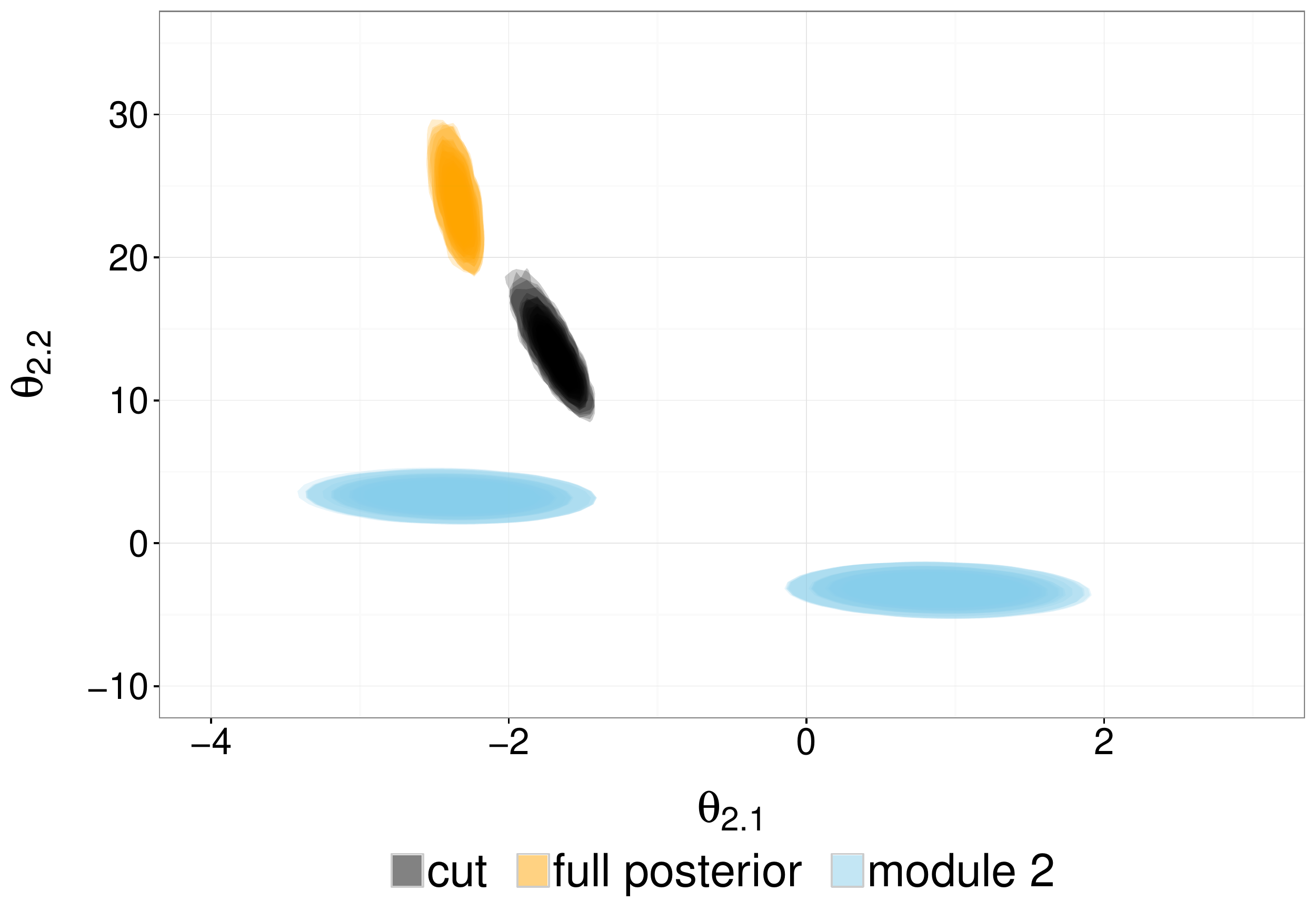}\includegraphics[width=0.5\textwidth]{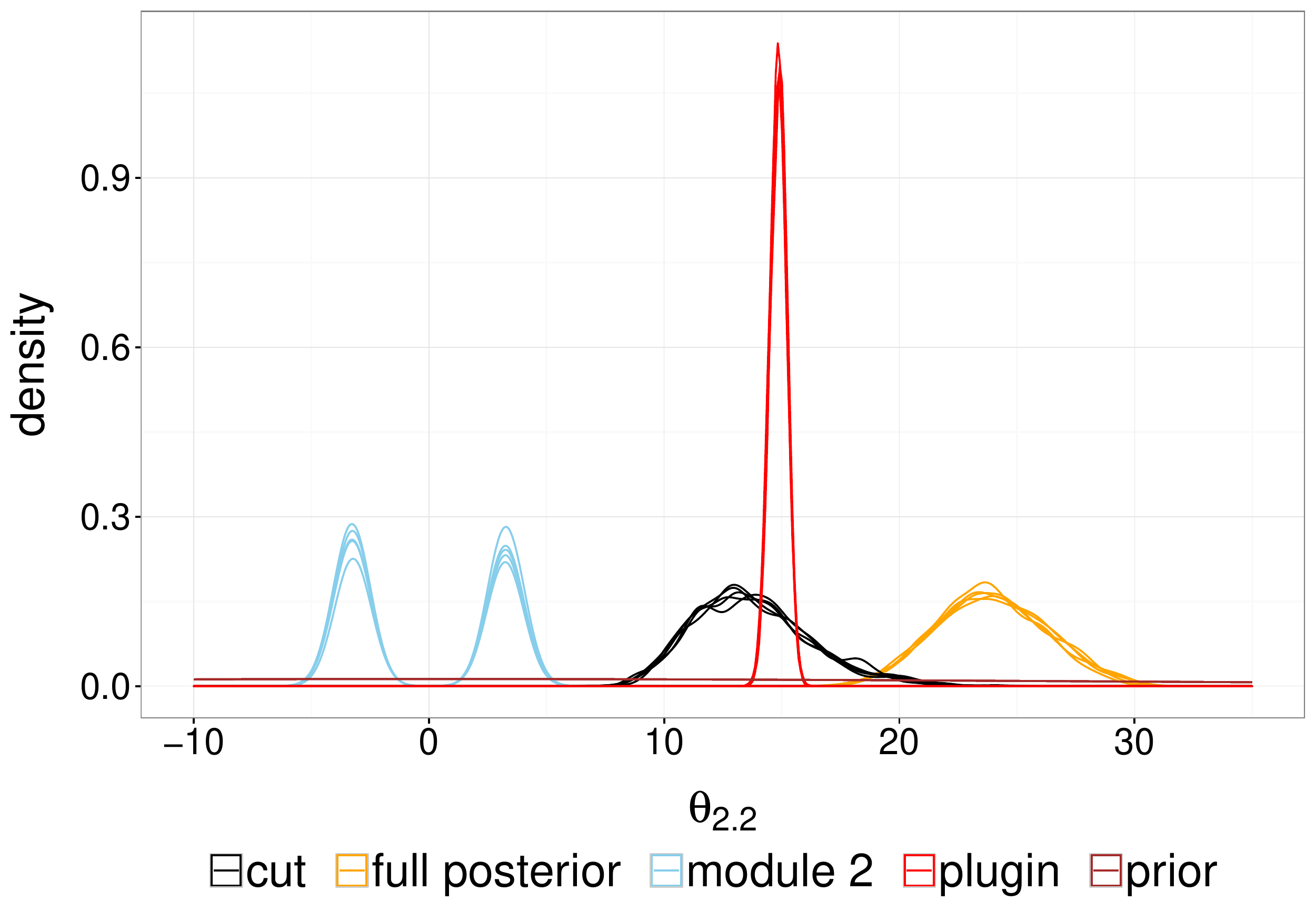}
\par\end{centering}
\caption{\label{fig:plummer:parameters2}Various candidate distributions for
$(\theta_{2,1},\theta_{2,2})$ (left) and marginal distribution of
$\theta_{2,2}$ (right), in the epidemiological study of Section \ref{sub:Epidemiological-study}.
The candidates overlap very little with one another. Based on the
predictive performance of $\theta_{1}$ in terms of predicting $Y_{1}$,
the cut distribution could be preferred, even though it yields a lower
predictive score for $Y_2$ than the full posterior or the posterior in the second
module only.}
\end{figure}

Some of the marginal distributions
of $\theta_{1}$ are shown in Figure \ref{fig:plummer:marginalparameters1}.
The full posterior is in agreement with the first module's posterior
for some parameters (such as $\theta_{1,1}$) but not for others (such
as $\theta_{1,9}$). The posterior in the second module is in disagreement
with the full posterior and the first module's posterior on most parameters.
We show the bivariate candidate distributions for
$(\theta_{2,1},\theta_{2,2})$, and the marginal distributions of
$\theta_{2,2}$ in Figure \ref{fig:plummer:parameters2}. We see that
the plug-in and the cut distributions give similar estimates for $\theta_{2,2}$,
but the cut distribution is more diffuse. Furthermore, it overlaps
very little with the full posterior distribution, so that decisions
derived from the cut approach would likely be different.

The predictive scores are given in Table \ref{tab:plummer}. If we
consider the task of predicting $Y_{1}$, we find that the full
model has worse predictive performance than the first module on its own,
but better predictive performance than the second module
on its own. This indicates that $Y_{2}$ does not help in predicting
$Y_{1}$. In this example, with only one observation per study,
the prior predictive performance in the first module corresponds
to the prequential predictive performance.  In terms of parameter estimation,
following our plan of action (Section \ref{sub:predictionplan}), we would use the first module on its own
to estimate $\theta_1$, and the cut distribution to estimate $(\theta_1,\theta_2)$,
since it is the only candidate considered with $\bar{\pi}(\theta_1|Y_1)$ as its first marginal.

If we consider the task of predicting $Y_{2}$, we find that
the full model has worse predictive performance than the second
module on its own. The cut approach yields a lower score, and
finally the plug-in approach yields the lowest score. As in the previous
section, the cut distribution is selected by our plan of action even though its
predictions for $Y_2$ yield a lower score than the predictions under the full posterior.

\begin{table}
\begin{centering}
\begin{minipage}[t]{0.4\columnwidth}%
    \input{plummer.predictY1}\end{minipage}\hspace{1cm}%
\begin{minipage}[t]{0.4\columnwidth}%
    \input{plummer.predictY2}\end{minipage}
\par\end{centering}
\caption{\label{tab:plummer}Predictive performance of various candidates in
the epidemiological study of Section \ref{sub:Epidemiological-study},
in terms of predicting $Y_{1}$ (left) and predicting $Y_{2}$ (right).
The numbers represent the average score over $5$ Monte Carlo runs,
with minimum and maximum values in brackets.}
\end{table}

\subsection{Propensity score\label{sub:Propensity-score}}

The propensity score methodology is used for causal inference in non-randomized
experiments \citep{rosenbaum1984reducing}. We will consider the setting of \citet{zigler2016central}
where the defects of the full posterior are explained in details. We use the example
to test whether the proposed procedure favors modular approaches in an automatic, data-driven way.

We consider the effect
of a variable $X$ (e.g. $X=1$ if ``exposure to a treatment'',
$X=0$ otherwise) on an outcome $Z$ (e.g. $Z=1$ if some event happens,
$Z=0$ otherwise), for a number of individuals $i\in\left\{ 1,\ldots,n\right\} $.
We have access to other covariates $C\in\mathbb{R}^{n\times p}$,
which might be correlated with both $X$ and $Z$ since the experiment
was not randomized. An attempt at correcting for confounding effects
goes as follows. First perform a logistic regression of $X$ on the
covariates $C$, 
\begin{equation}
\forall i\in\{1,\ldots,n\}\quad\text{logit}\,\mathbb{P}\left(X_{i}=1\mid C_{i}\right)=\theta_{1,0}+\sum_{j=1}^{p}\theta_{1,j}C_{ij},\label{eq:propensity:model}
\end{equation}
where $C_{i}$ denotes the $i$-th row of $C$. The quantity $e_{i}=\mathbb{P}\left(X_{i}=1\mid C_{i}\right)$
is referred to as the propensity score of individual $i$, and is
a scalar summary of the relationship between the covariates and the
treatment variable. For our purposes, the above logistic regression
defines a first module with parameters $\theta_{1}=(\theta_{1,0},\ldots,\theta_{1,p})$,
on which a centered Normal prior distribution is specified. The prior
variance is set to $800$ on the intercept and to $50$ for the other
coefficients.

One can proceed to the regression of $Z$ on $X$ over groups of individuals
that share similar propensity scores. We consider a stratification
of the scores $e_{i}$ in quintiles; the variable $q_{i}\in\left\{ 1,\ldots,5\right\} $
for each $i\in\{1,\ldots,n\}$ indicates to which quintile each $e_{i}$
belongs. The vector $q$ is deterministic given $e$, and thus deterministic
given $\theta_{1}$, $X$ and $C$. The effect of $X$ on $Z$ can
be modelled with another a logistic regression, 
\begin{equation}
\forall i\in\{1,\ldots,n\}\quad\text{logit}\,\mathbb{P}\left(Z_{i}=1\mid X_{i},q_{i}\right)=\theta_{2,0}+\theta_{2,1}X_{i}+\sum_{k=2}^{5}\theta_{2,k}\delta_{k}\left(q_{i}\right),\label{eq:outcome:model}
\end{equation}
where $\delta_{k}(q_{i})$ is equal to $1$ if $q_{i}=k$, and $0$
otherwise. Along with a Normal prior on $\theta_{2}=(\theta_{2,0},\ldots,\theta_{2,5})$,
centered at zero and with variance $800$ on the intercept and $50$
on the other coefficients, this concludes the specification of the
second module. The object of interest might be the parameter $\theta_{2,1}$
(or by-products of it), which is the coefficient of the treatment
variable in the above logistic regression.

Standard practice in this setting is to obtain the propensity scores
$e_{i}$ from the first module only, and to plug them into the second module
to estimate $\theta_{2,1}$. Indeed, the goal of the propensity score
approach is to compensate for lack of randomization in the experiment.
In a randomized experiment, the assignment of $X$ is, by design, independent of the  
covariates $C$, and the outcome $Z$ would not be
observed at the time when $X$ is assigned. Therefore, it seems odd
to use the outcome variable $Z$ in order to estimate propensity scores
that relate to the treatment assignment part of the problem. However,
if we do have access to $Z$ from the onset, it is legitimate to wonder
whether it should be used in the estimation of propensity scores.
A series of interesting articles \citep{Zigler:Watts:2013,zigler2014uncertainty,zigler2016central}
investigates this question. We consider the experiment described in
\citet{zigler2016central}, where $n=1000$ individuals have $p=6$
covariates, generated as independent Normal realizations: $C_{ij}\sim\mathcal{N}(0,1)$ for
all $i,j$. The treatment
variable $X_{i}\in\{0,1\}$ is generated according to a logistic regression:
\[
\forall i\in\{1,\ldots,n\}\quad\text{logit}\,\mathbb{P}\left(X_{i}=1\mid C_{i}\right)=\theta_{1,0}^{\star}+\sum_{j=1}^{6}\theta_{1,j}^{\star}C_{ij},
\]
where $\theta_{1}^{\star}=(0,0.1,0.2,0.3,0.4,0.5,0.6)$: the first
module is well-specified. Given $X$ and $C$, we generate an outcome
variable $Z$ via the equation 
\begin{align*}
\forall i\in\{1,\ldots,n\}\quad\text{logit}\mathbb{P}\left(Z_{i}=1\mid X_{i},C_{i}\right)= & \gamma_{1}^{\star}C_{i1}+\gamma_{2}^{\star}\exp\left(C_{i2}-1\right)\\
 & +\gamma_{3}^{\star}C_{i3}+\gamma_{4}^{\star}\exp\left(C_{i4}-1\right)+\gamma_{5}^{\star}|C_{i5}|+\gamma_{6}^{\star}|C_{i6}|,
\end{align*}
where $\gamma^{\star}=(0.6,0.5,0.4,0.3,0.2,0.1)$: the second module
is misspecified. One question is whether the model can be used to
investigate causal effects of $X$ on $Z$, despite that misspecification.
Here, given the covariates $C$, the outcome $Z$ is generated independently
of $X$, so that we hope that a statistical approach to causal inference
would conclude at an absence of causal effect of $X$ on $Z$: in
other words, we want to estimate $\theta_{2,1}$ close to zero.

\begin{figure}
\begin{centering}
    \includegraphics[width=0.3\textwidth]{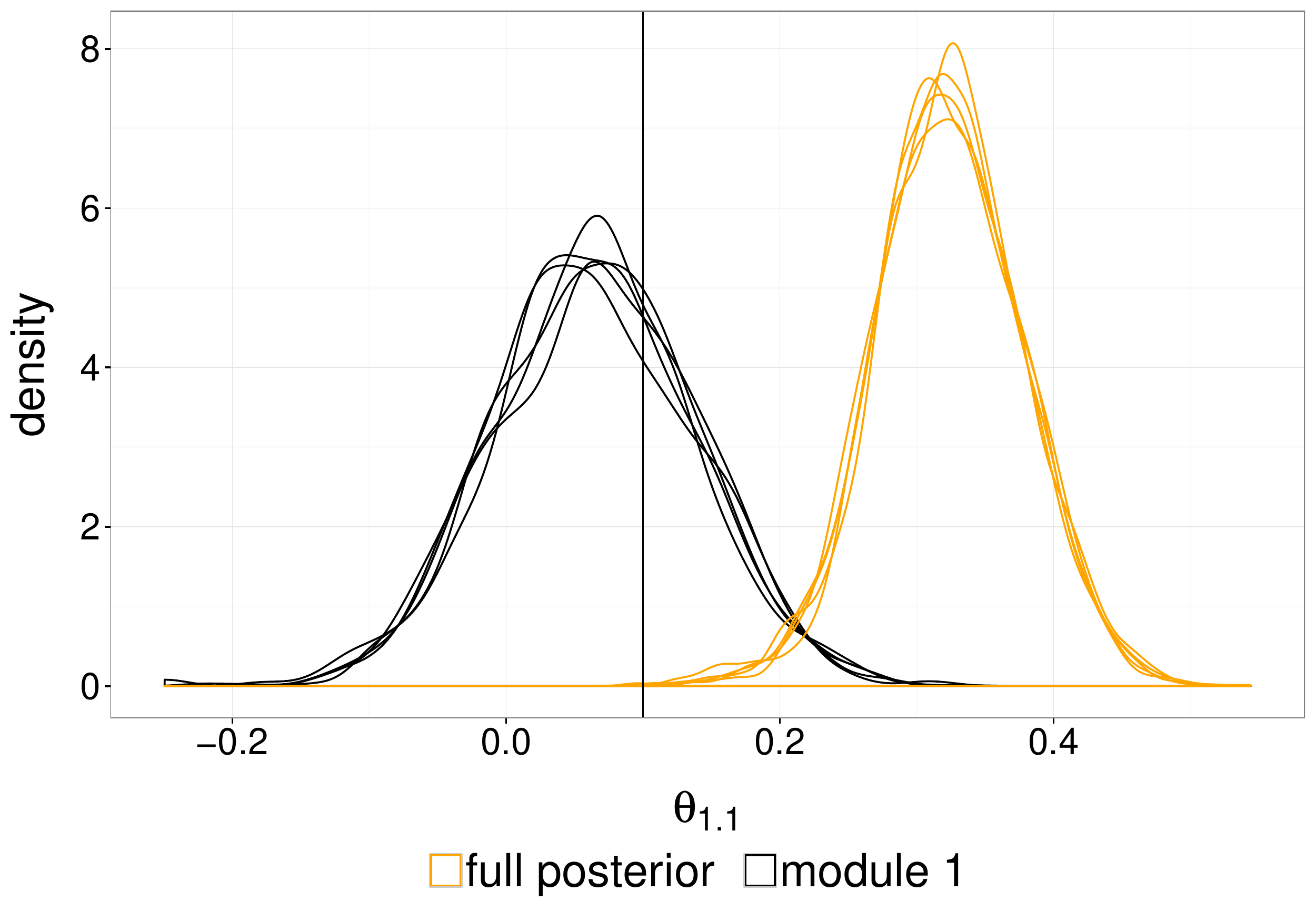}\includegraphics[width=0.3\textwidth]{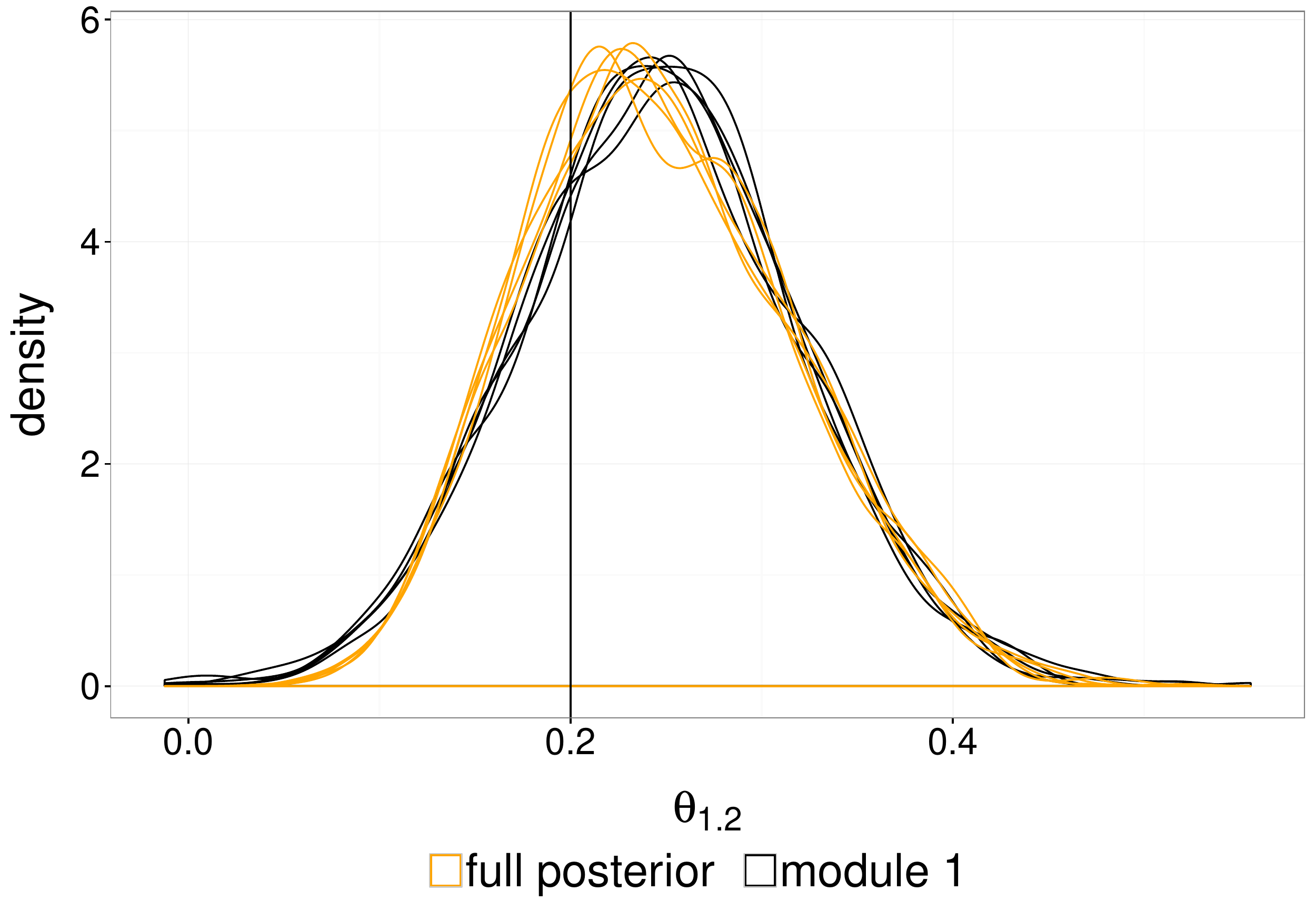}
    \includegraphics[width=0.3\textwidth]{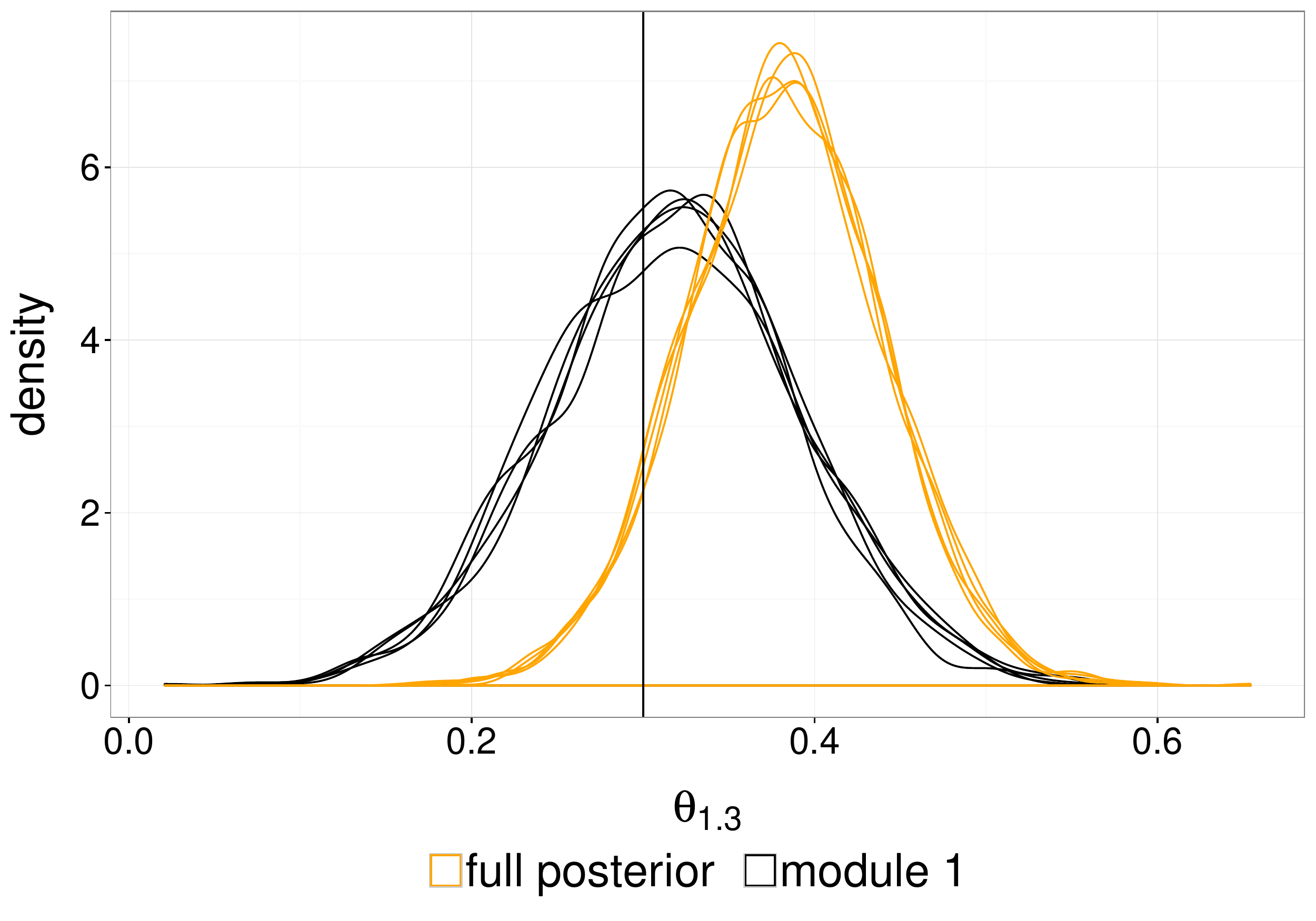}
\par\end{centering}
\caption{\label{fig:propensity:marginal1} Distribution of some regression
coefficients in the first module, in the propensity score example
of Section \ref{sub:Propensity-score}. The values used to generate
the data are indicated by vertical black lines.}
\end{figure}

Some marginal distributions of $\theta_{1}$
are shown in Figure \ref{fig:propensity:marginal1}. The posterior under
the second module only yields a very flat distribution on the regression
parameters, omitted from the plots for clarity. As in Section \ref{sub:Epidemiological-study},
we would not expect users to consider the posterior given $Z$
alone, as the propensity scores would then result from prior information
only, but we include the associated score in the tables, to emphasize
that it would yield the best predictive performance for $Z$. In Figure
\ref{fig:propensity:marginal1} we see that the first posterior puts
its mass near the data-generating values, whereas the full posterior
is sometimes in disagreement, e.g. for $\theta_{1,1}$.

\begin{figure}
\begin{centering}
    \includegraphics[width=0.5\textwidth]{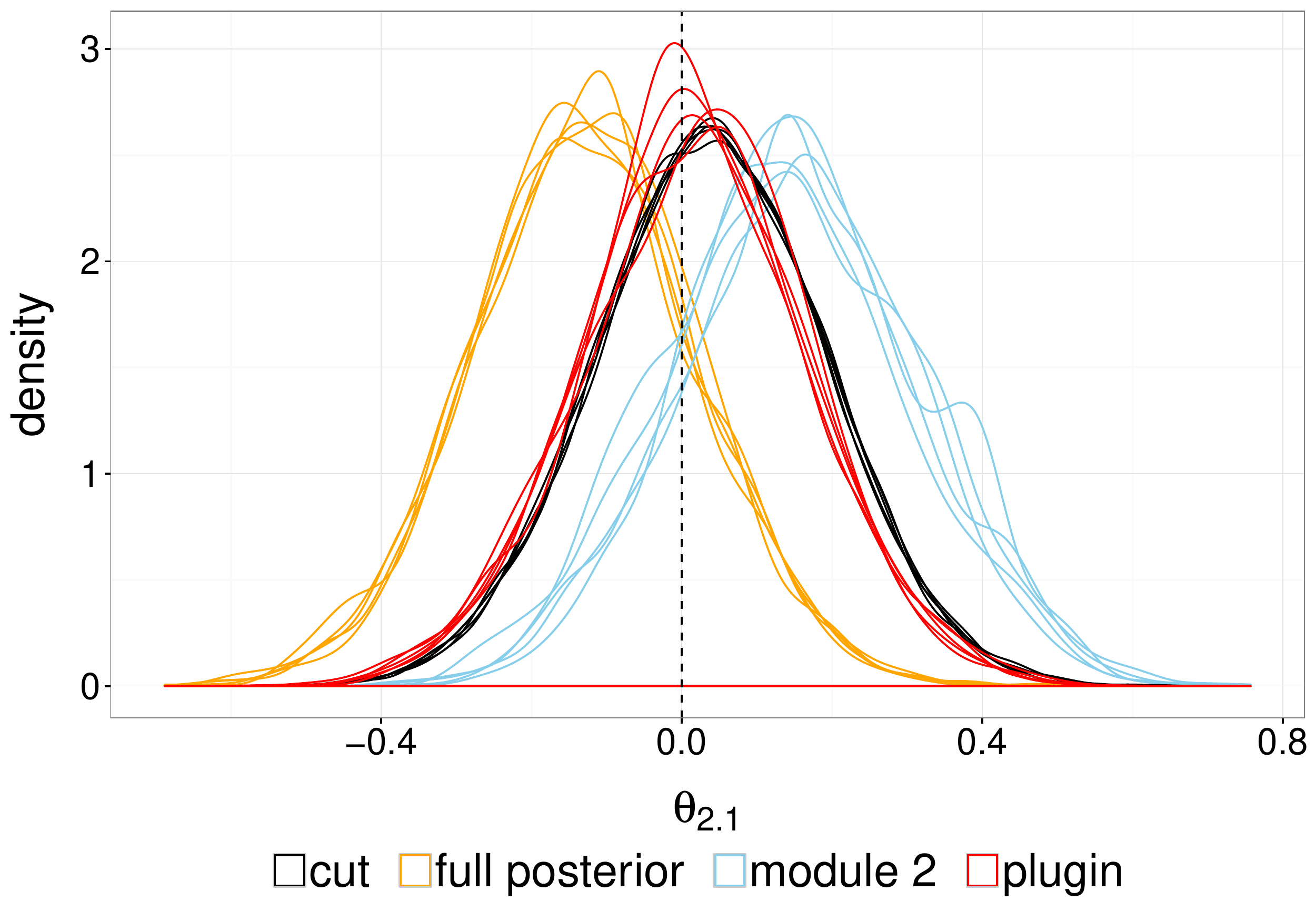}
\par\end{centering}
\caption{\label{fig:propensity:marginal2} Distribution of $\theta_{2,1}$,
the coefficient of $X$ in the regression of $Z$ on $X$ and the
propensity score quintile indicators, in the propensity score example
of Section \ref{sub:Propensity-score}. The vertical dashed line at
zero indicates that we expect to retrieve zero as the result of the
inference, since the values of $Z$ are generated irrespectively of
the values of $X$.}
\end{figure}

Some candidates distributions for $\theta_{2,1}$ are displayed in Figure \ref{fig:propensity:marginal2}.
The absence of causal effect of $X$ on $Z$ can indeed be retrieved
from either the cut and the plug-in distributions, the expectations of which are
close to zero. The full posterior is shifted towards negative values,
while the posterior in the second module puts more mass on positive
values. 

Table \ref{tab:propensity} contains the predictive scores. For this example, our proposed
plan of action again leads to modular approaches over the full posterior. For $\theta_1$, looking 
at the predictive performance for $Y_1$, we would use the
first module posterior. To preserve $\pi_1(\theta_1|X,C)$
as the first marginal of a distribution on $(\theta_1,\theta_2)$, we would choose the cut distribution, 
even though it yields lower predictive scores for $Y_2$.
This experiment illustrates again that modular approaches
favored by practitioners can be validated by quantitative criteria,
and that the full posterior can underperform in the presence of misspecification.

\begin{table}
\begin{centering}
\begin{minipage}[t]{0.4\columnwidth}%
\input{propensity.predictY1}\end{minipage}\hspace{1cm}%
\begin{minipage}[t]{0.4\columnwidth}%
\input{propensity.predictY2}\end{minipage}
\par\end{centering}
\caption{\label{tab:propensity}Predictive performance of various candidates
in the propensity score example of Section \ref{sub:Propensity-score},
in terms of predicting $X$ in the first module (left) and predicting
$Z$ in the second one (right). The numbers represent the average
score over $5$ Monte Carlo runs, with minimum and maximum values
in brackets.}
\end{table}

\subsection{Meta-analysis \label{sec:metaanalysis}}

Here we go beyond models made of two modules, 
to illustrate how the proposed procedure can be adapted in other settings.
Calculations are provided in Appendix \ref{app:meta:calculations}.

\subsubsection{Model and candidate distributions}

This example is again taken from \citet{liu2009}, where it raises concerns
about the full posterior in certain hierarchical models. We use the example
to test whether the defects of the full posterior can be detected automatically
with the proposed approach.
Consider $N$ studies, indexed by $i\in 1:N$, and $n_{i}$
individuals in each study $i$, indexed by $j\in1:n_{i}$. The entire
data set is denoted by $y=(y_{i,j})$, and the data of study $i$
by $y_{i}=(y_{i,1},\ldots,y_{i,n_{i}})$. The average of the observations
in study $i$ is $\bar{y}_{i}=n_{i}^{-1}\sum_{j=1}^{n_{i}}y_{i,j}$
and we also introduce $s_{i}^{2}=\sum_{j=1}^{n_{i}}\left(y_{i,j}-\bar{y}_{i}\right)^{2}$.
The model specifies, for all $i\in1:N$ and all $j\in1:n_{i}$, that
the observation $y_{i,j}$ follows the distribution $\mathcal{N}(b_{i},\sigma_{i}^{2})$,
where $\sigma_i^2$ here denotes the variance. 
The prior distribution on $(b_{i},\sigma_{i}^{2})$ for all $i\in1:N$
and on $\tau^{2}$ is specified as 
\begin{equation}
b_{i}\sim\mathcal{N}(0,\tau^{2}),\quad p(\sigma_{i}^{2})\propto\frac{1}{\sigma_{i}^{2}},\quad p(\tau^{2}|\sigma_{1:N}^{2})\propto\frac{1}{\tau^{2}+N^{-1}\sum_{i=1}^{N}(\sigma_{i}^{2}/n_{i})}.\label{eq:prior}
\end{equation}
The above is a reference prior according to \citet{liu2009}. The link between
the studies is the variance parameter $\tau^{2}$. The posterior distribution
of $\sigma_{i}^{2}$ given $\tau^{2}$ in study $i\in1:N$, integrating
$b_{i}$ out, is given by 
\begin{equation}
p(\sigma_{i}^{2}|\tau^{2},y_{i})\propto\frac{\sigma_{i}^{-n_{i}-1}}{\sqrt{\tau^{2}+\sigma_{i}^{2}/n_{i}}}\exp\left(-\frac{1}{2}\frac{\left(\bar{y}_{i}\right)^{2}}{\left(\tau^{2}+\sigma_{i}^{2}/n_{i}\right)}-\frac{s_{i}^{2}}{2\sigma_{i}^{2}}\right).\label{eq:posteriorsigma_i}
\end{equation}
This leads to the full posterior of $\tau^{2},\sigma_{1:N}^{2}$ given
$y$:
\begin{align}
p(\tau^{2},\sigma_{1:N}^{2}|y) & \propto\frac{1}{\tau^{2}+N^{-1}\sum_{i=1}^{N}(\sigma_{i}^{2}/n_{i})}\prod_{i=1}^{N}\frac{\sigma_{i}^{-n_{i}-1}}{\sqrt{\tau^{2}+\sigma_{i}^{2}/n_{i}}}\exp\left(-\frac{1}{2}\frac{\left(\bar{y}_{i}\right)^{2}}{\left(\tau^{2}+\sigma_{i}^{2}/n_{i}\right)}-\frac{s_{i}^{2}}{2\sigma_{i}^{2}}\right).\label{eq:posteriortausigma}
\end{align}
We can evaluate the above expression for all non-negative values of
$\tau^{2},\sigma_{1:N}^{2}$, and thus we can perform Markov chain Monte Carlo (MCMC) to approximate
the full posterior $p(\tau^{2},\sigma_{1:N}^{2}|y)$. Finally, given
$(\tau^{2},\sigma_{1:N}^{2},y)$, the conditional distribution of
$b_{i}$ is given by:
\begin{equation}
b_{i}\sim\mathcal{N}\left(\frac{\bar{y}_{i}\tau^{2}}{\tau^{2}+\sigma_{i}^{2}/n_{i}},\frac{\tau^{2}\sigma_{i}^{2}/n_{i}}{\tau^{2}+\sigma_{i}^{2}/n_{i}}\right).\label{eq:bsgivenrest}
\end{equation}

In \citet{liu2009}, the cut distribution is introduced as follows. The conditional distribution
of $\tau^{2}$ given $(\sigma_{1:N}^{2},y)$ is that of the full posterior
(dropping constants in $\tau^{2})$:
\begin{equation}
p(\tau^{2}|\sigma_{1:N}^{2},y)\propto\frac{1}{\tau^{2}+N^{-1}\sum_{i=1}^{N}(\sigma_{i}^{2}/n_{i})}\prod_{i=1}^{N}\frac{\sigma_{i}^{-n_{i}-1}}{\sqrt{\tau^{2}+\sigma_{i}^{2}/n_{i}}}\exp\left(-\frac{1}{2}\frac{\left(\bar{y}_{i}\right)^{2}}{\left(\tau^{2}+\sigma_{i}^{2}/n_{i}\right)}\right),\label{eq:posteriortaugivensigmas}
\end{equation}
with an intractable normalizing constant that is a function of $\sigma_{1:N}^{2}$.
The marginal distribution of $\sigma_{1:N}^{2}$ given $y$ is specified
as 
\begin{equation}
p^{\text{cut}}(\sigma_{1:N}^{2}|y)=\prod_{i=1}^{N}p(\sigma_{i}^{2}|y_{i})\propto\prod_{i=1}^{N}\sigma_{i}^{-n_{i}-1}\exp\left(-\frac{s_{i}^{2}}{2\sigma_{i}^{2}}\right),\label{eq:marginalsigmacut}
\end{equation}
which is a product of Inverse Gamma densities. This is defined as
long as $n_{i}\geq2$ for all $i\in1:N$. The joint cut distribution
of $(\tau^{2},\sigma_{1:N}^{2})$ is the product of the marginal of
Eq. (\ref{eq:marginalsigmacut}) and the conditional of Eq. (\ref{eq:posteriortaugivensigmas}):
\begin{equation}
p^{\text{cut}}(\tau^{2},\sigma_{1:N}^{2})=p(\tau^{2}|y,\sigma_{1:N}^{2})p^{\text{cut}}(\sigma_{1:N}^{2}|y).\label{eq:meta:jointcut}
\end{equation}
Because of the intractable normalizing constant in $p(\tau^{2}|y,\sigma_{1:N}^{2})$,
we cannot directly perform MCMC to approximate $p^{\text{cut}}(\tau^{2},\sigma_{1:N}^{2})$.
The constant could be accurately approximated, since it is the integral
of the right-hand side of Eq. (\ref{eq:posteriortaugivensigmas})
with respect to $\tau^{2}$, which is a one-dimensional variable.
We will prefer the following procedure that generates i.i.d. draws
from the cut distribution.

We can obtain i.i.d. samples from $p^{\text{cut}}(\sigma_{1:N}^{2}|y)$
by inverting Gamma variables, and then sample $\tau^{2}$ given such
draws, using rejection sampling. To this aim, we perform a reparametrization
of $\tau^{2}$, introducing $u=(1+\tau^{2})^{-1}$, with inverse $\tau^{2}=u^{-1}-1$.
Applying that change of variable to Eq. (\ref{eq:posteriortaugivensigmas}),
the distribution of $u$ given $(\sigma_{1:N}^{2},y)$ has density proportional to
\[
\frac{1}{u^{-1}-1+N^{-1}\sum_{i=1}^{N}(\sigma_{i}^{2}/n_{i})}\prod_{i=1}^{N}\frac{\sigma_{i}^{-n_{i}-1}}{\sqrt{u^{-1}-1+\sigma_{i}^{2}/n_{i}}}\exp\left(-\frac{1}{2}\frac{\left(\bar{y}_{i}\right)^{2}}{\left(u^{-1}-1+\sigma_{i}^{2}/n_{i}\right)}\right)\times\left|\frac{1}{u^{2}}\right|.
\]
The maximum value of the above expression,
over all $u\in[0,1]$, can be obtained numerically, which enables
rejection sampling; we will use a Uniform distribution on $[0,1]$
as a proposal distribution. 

\subsubsection{The issue with the full posterior \label{sec:metaanalysis:issue}}

We illustrate the issue with the full posterior and
the appeal of the cut distribution in a numerical experiment 
inspired by the discussion in \citet{liu2009}
about the defects of the full posterior distribution
in misspecified random-effects models. 

\begin{figure}
    \begin{centering}
        \includegraphics[width=0.85\textwidth]{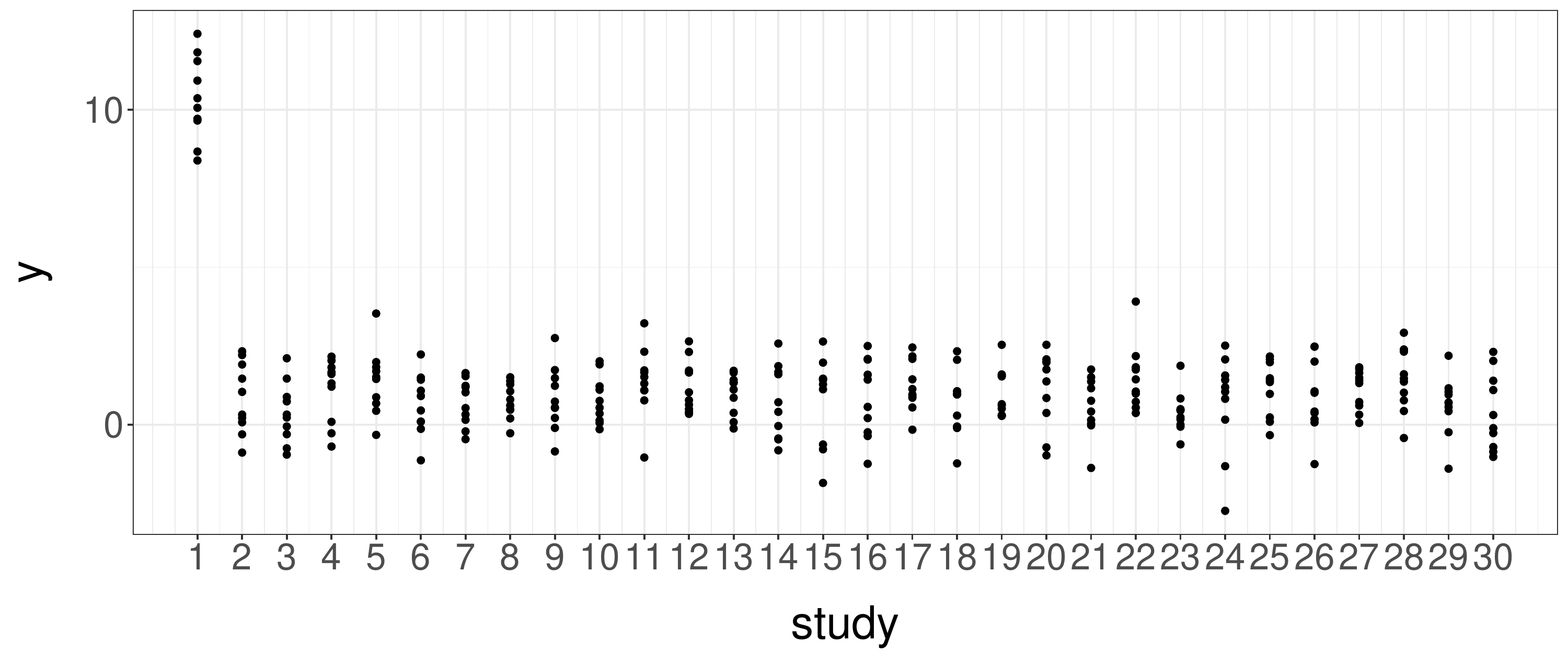}
        \par\end{centering}
    \caption{Data generated from the meta-analysis model of Section \ref{sec:metaanalysis}, with $b_{1}^{\star}=10$,
    $b_{i}^{\star}=1$ for $i\in2:N$ and $\sigma_{i}^{\star}=1$ for
    $i\in1:N$, with $N=30$ studies and $n_i = 10$ individuals per study. \label{fig:metanalysis:data} }
\end{figure}

\begin{figure}
    \centering
    \begin{subfigure}[t]{0.70\textwidth}
        \centering
        \includegraphics[width=\textwidth]{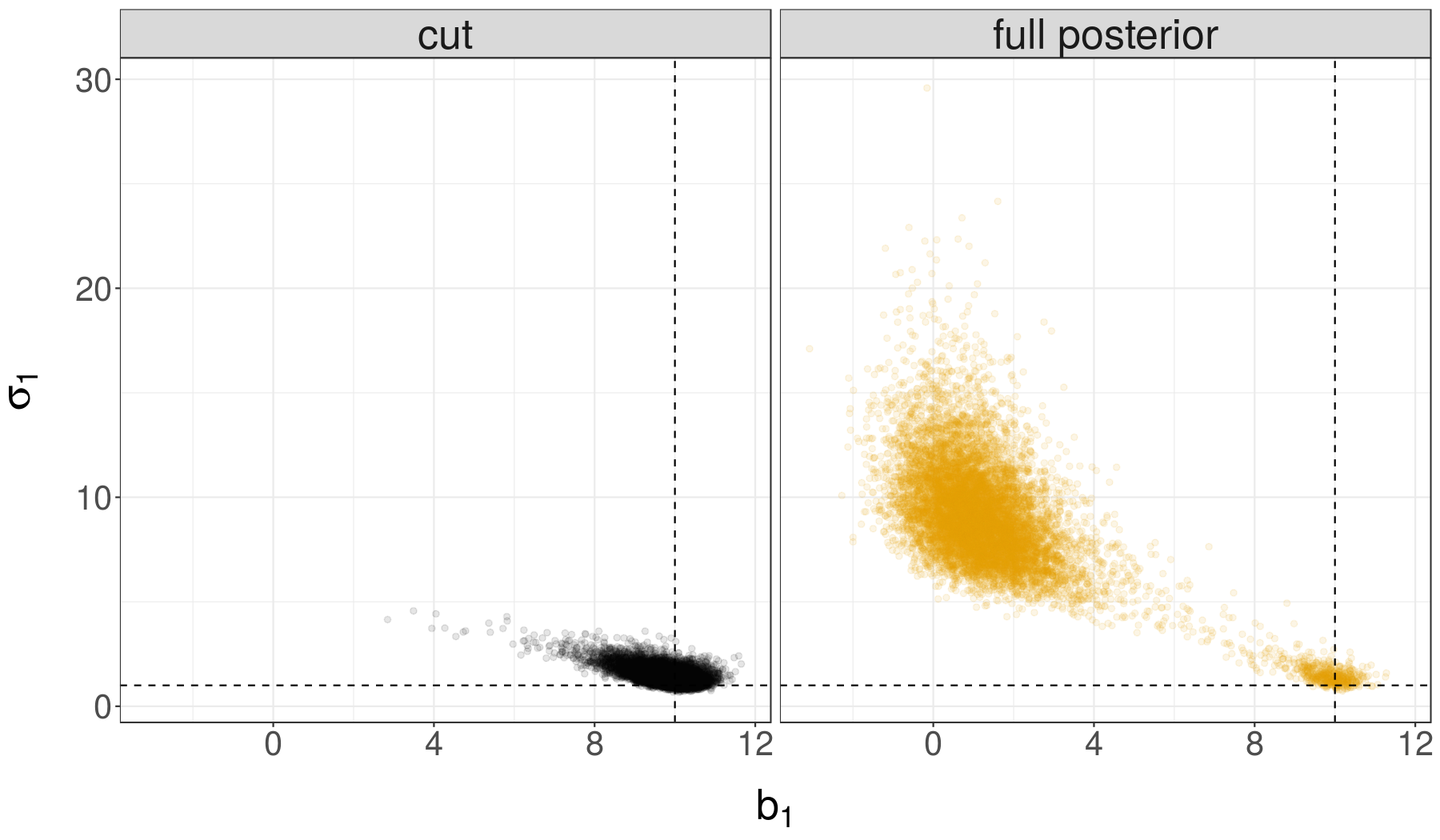}
        \caption{Marginal distribution of $(b_{1},\sigma_{1})$.}
    \end{subfigure}

    \begin{subfigure}[t]{0.70\textwidth}
        \centering
        \includegraphics[width=\textwidth]{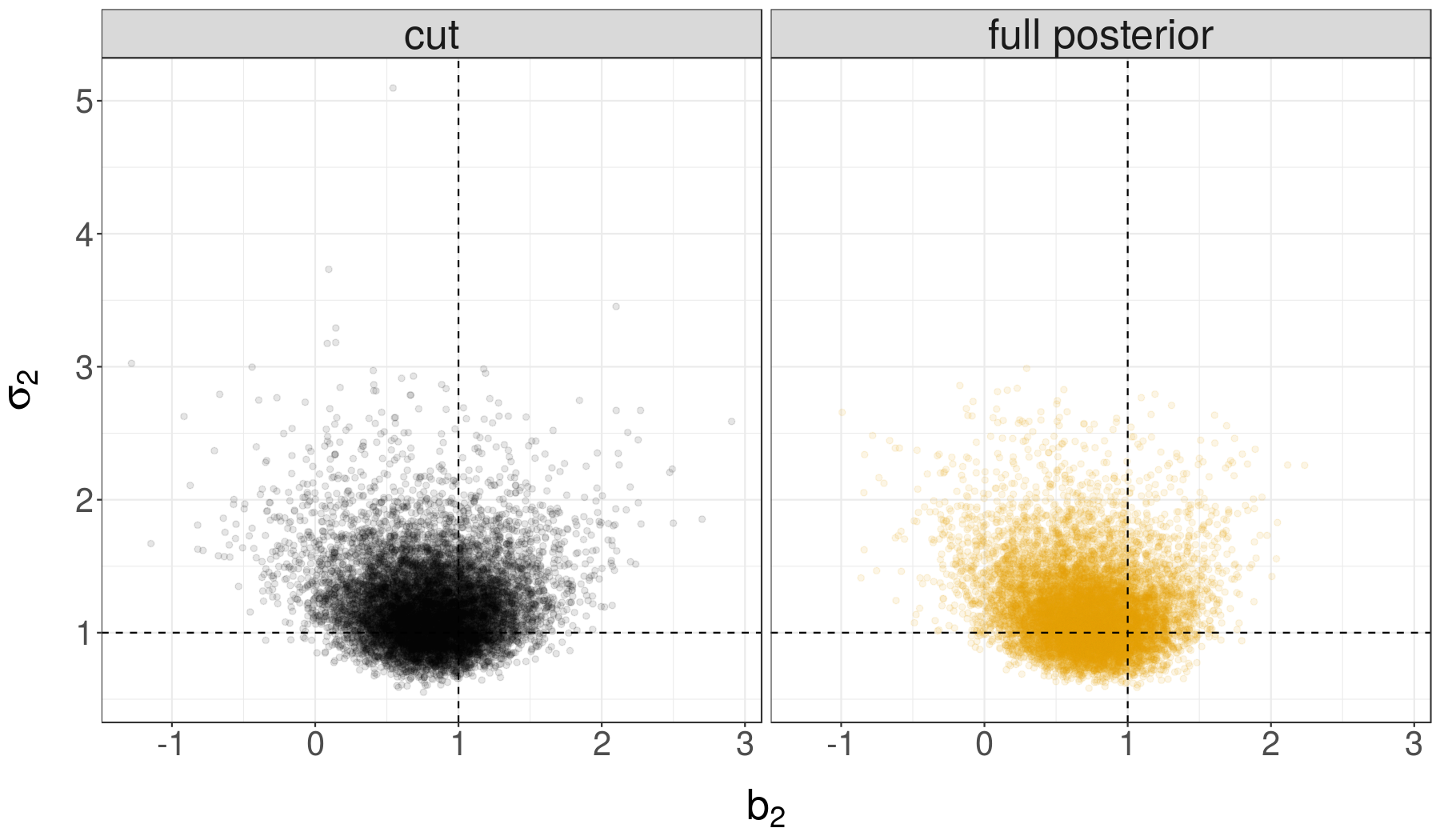}
        \caption{Marginal distribution of $(b_{2},\sigma_{2})$.}
    \end{subfigure}

    \begin{subfigure}[t]{0.70\textwidth}
        \centering
        \includegraphics[width=\textwidth]{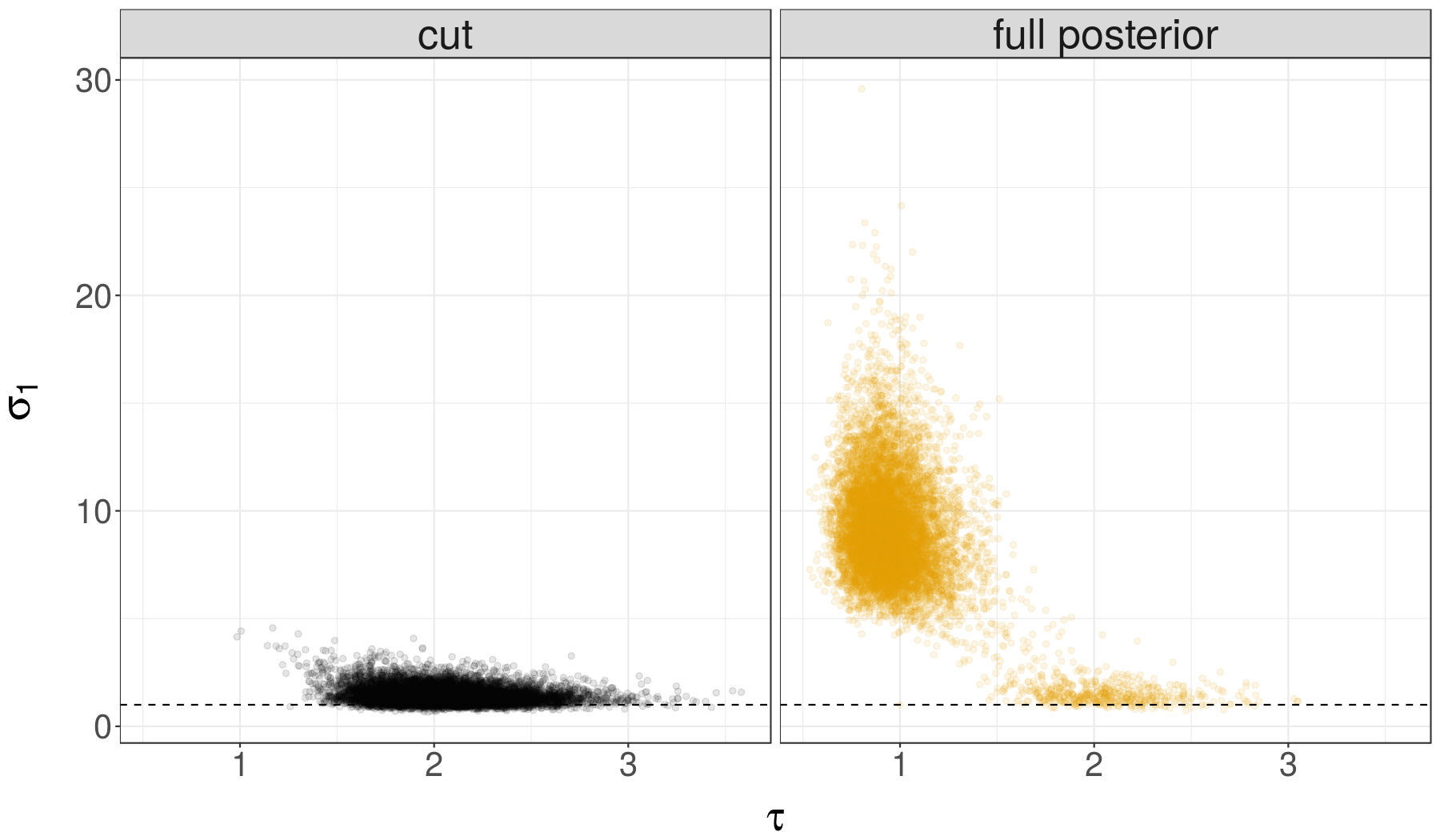}
        \caption{Marginal distribution of $(\tau,\sigma_{1})$.}
    \end{subfigure}
    \caption{Marginal posterior distributions of $(b_{1},\sigma_{1})$, $(b_{2},\sigma_{2})$
    (top and middle row), and of $(\tau,\sigma_{1})$ (bottom row) under the full
model posterior and the cut distribution in the meta-analysis example of Section \ref{sec:metaanalysis}. The distributions of $(b_{i},\sigma_{i})$ for $i\in 3:n$ are similar
    to that of $(b_{2},\sigma_{2})$, and thus not shown. \label{fig:metanalysis:marginalsbisgma} }
\end{figure}

We set $N=30$ and $n_{i}=10$ for all $i\in1:N$.
The data-generating parameters are set to $b_{1}^{\star}=10$, $b_{i}^{\star}=1$
for $i\in2:N$, and $\sigma_{i}^{\star}=1$ for all $i\in1:N$. The
data are shown in Figure \ref{fig:metanalysis:data}. 
We obtain the parameters $(\tau,\sigma_{1:N},b_{1:N})$ given the
data under both the full posterior distribution and the cut approach.
The marginal distributions of $(\sigma_{1},b_{1})$ and $(\sigma_{2},b_{2})$
are shown in Figure \ref{fig:metanalysis:marginalsbisgma}. The marginals
of $(\sigma_{i},b_{i})$ for $i\in3:N$ are similar to that of $(\sigma_{2},b_{2})$,
and are thus not shown. The horizontal and vertical dashed lines indicate
the values of $b_{i}^{\star}$ and $\sigma_{i}^{\star}$. From the
plots, it is apparent that the distribution obtained with the cut
approach is more able to retrieve the data-generating values. Note
the bimodality of the full posterior distribution of $(\sigma_{1},b_{1})$,
with a minor mode located where the cut distribution puts most of
its mass. The marginal distribution of $(\tau,\sigma_{1})$ is also
shown, and also features two modes.
From the numerical experiments, the cut distribution
puts most of its mass close to the data-generating
values $b_{1}^{\star},\sigma_{1}^{\star}$, whereas the full posterior
distribution does not.

\subsubsection{Predictive criteria}

We are now interested in a data-driven predictive criterion to decide
between the full posterior and the cut approach, without knowledge
of the data-generating values $b_{1:N}^{\star},\sigma_{1:N}^{\star}$.
In the present context, several obstacles appear in the way of our
proposed plan of action.
\begin{itemize}
    \item[-] The number of observations, $n_{i}$ in study $i$, enters the specification
        of the prior distribution. Therefore, the task of sequential prediction
        is ambiguous: either we specify the prior once and for all, using
        $n_{i}$ in study $i$, or we redefine the prior sequentially as we
        assimilate more and more observations, replacing $n_{i}$ in the prior
        by the current number of observations. We choose to fix $n_{i}$ in
        the prior specification, even when we predict $y_{i,j}$ given $y_{i,1:j-1}$
        for $j<n_{i}$.
    \item[-] The prior is improper, and the cut distribution is well-defined only
        if $n_{i}\geq2$ for any $i\in1:N$.
    \item[-] We can either define a single criterion quantifying the quality of
        the predictions of all observations, or we can define a criterion
        for each study, quantifying the predictive quality conditional upon
        the data of the other studies. We choose the latter, enabling
        the identification of problematic studies.
\end{itemize}
We now describe the proposed study-specific predictive criterion. For a
study $i\in1:N$, denote by $y_{\setminus i}$ the data of the other
studies, and $\sigma_{\setminus i}^{2}$ denotes $\sigma_{i^{\prime}}^{2}$
for all $i^{\prime}\neq i$. For the task of predicting the
observations $y_{i,j}$, for $j\in1:n_{i}$, given $y_{\setminus i}$
and $y_{i,1},\ldots,y_{i,j-1}$, we will condition on the first two
observations, $y_{i,1},y_{i,2}$, otherwise the cut distribution in
Eq. \eqref{eq:meta:jointcut} would not be defined. 
We introduce a predictive
criterion under the full posterior approach before introducing a comparable
criterion for the modularized approach. 

First, we obtain a sample approximating the distribution of $(\tau^{2},\sigma_{1:N}^{2})$
given $(y_{\setminus i},y_{i,1},\ldots,y_{i,j})$, as in Eq. (\ref{eq:posteriortausigma}),
using only $j$ observations, $y_{i,1:j}$, in the $i$-th study;
we still use $n_{i}$ in the definition of the prior distribution
of $\tau^{2}$ conditional on $\sigma_{1:N}^{2}$. We then proceed
to computing, for all $j\in\{3,\ldots,n_{i}\}$,
\begin{equation}
p(y_{i,j}|y_{\setminus i},y_{i,1},\ldots,y_{i,j-1})=\int p(y_{i,j}|\sigma_{i}^{2},\tau^{2},y_{i,1},\ldots,y_{i,j-1})p(\tau^{2},\sigma_{\setminus i}^{2},\sigma_{i}^{2}|y_{\setminus i},y_{i,1},\ldots,y_{i,j-1})d\tau^{2}d\sigma_{1:N}^{2},\label{eq:predictivenext:std}
\end{equation}
where we can approximate the integral by Monte Carlo samples. Note
that we can either do this for each $j\in\{3,\ldots,n_{i}\}$, independently,
or we can use a sequential Monte Carlo sampler, starting from the
distribution of $(\tau^{2},\sigma_{1:N}^{2})$ given $(y_{\setminus i},y_{i,1:2})$
and assimilating $y_{i,3:n_{i}}$ one by one. We then aggregate the
predictive scores of each observation, leading to the criterion 
\[
\log p(y_{i,3:n_{i}}|y_{\setminus i},y_{i,1:2})=\sum_{j=3}^{n_{i}}\log p(y_{i,j}|y_{\setminus i},y_{i,1},\ldots,y_{i,j-1}),
\]
which is quantity that would also appear in a partial Bayes factor \citep{o1995fractional,berger1996intrinsic}.
Since the ordering of the observations in each study is arbitrary,
we could average the above criterion over the ``2 choose $n_{i}$''
choices of first two observations, at the cost of more calculations.

Under the cut approach, we define similarly, for all $j\in\{3,\ldots,n_{i}\}$,
\begin{align*}
p^{\text{cut}}(y_{i,j}|y_{\setminus i},y_{i,1},\ldots,y_{i,j-1}) & =\int p(y_{i,j}|\sigma_{i}^{2},\tau^{2},y_{i,1},\ldots,y_{i,j-1})p^{\text{cut}}(\tau^{2},\sigma_{\setminus i}^{2},\sigma_{i}^{2}|y_{\setminus i},y_{i,1},\ldots,y_{i,j-1})d\tau^{2}d\sigma_{1:N}^{2},\\
\log p^{\text{cut}}(y_{i,3:n_{i}}|y_{\setminus i},y_{i,1:2}) & =\sum_{j=3}^{n_{i}}\log p^{\text{cut}}(y_{i,j}|y_{\setminus i},y_{i,1},\ldots,y_{i,j-1}),
\end{align*}
and again, we could average over permutations of $y_{i,1:n_{i}}$
at the cost of extra calculations.

\begin{figure}
    \centering
    \includegraphics[width=0.85\textwidth]{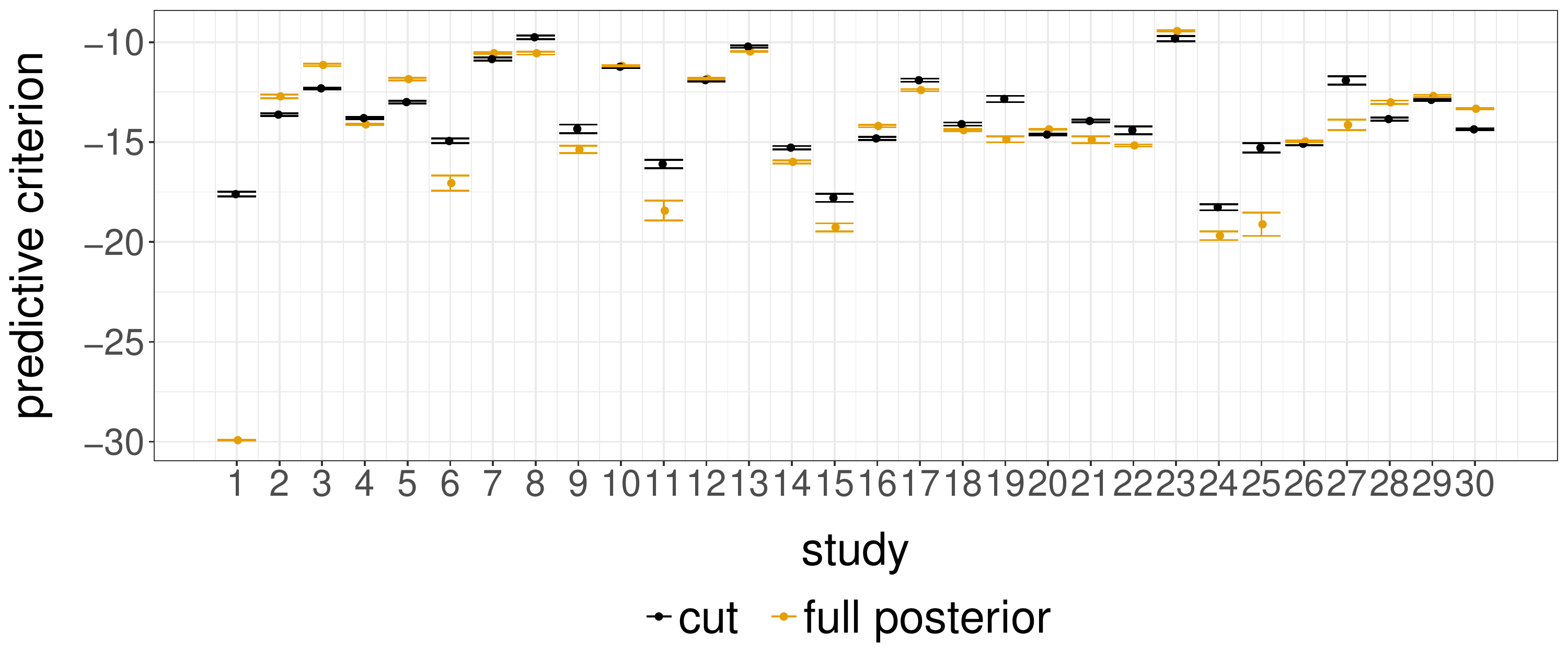}
    \caption{Prediction criterion for each study given the others, in the meta-analysis
        model, under the cut approach and the full posterior. The error bars
        show the mean plus and minus two standard errors, of criteria computed
        $10$ times independently, using Monte Carlo methods. \label{fig:metanalysis:prediction} }
\end{figure}

The two quantities $\log p(y_{i,3:n_{i}}|y_{\setminus i},y_{i,1:2})$
and $\log p^{\text{cut}}(y_{i,3:n_{i}}|y_{\setminus i},y_{i,1:2})$
can then be compared for each study $i$. Indeed, they monitor the
performance of sequential probabilistic predictions of the same observations,
under the same logarithmic scoring rule. Figure \ref{fig:metanalysis:prediction}
shows the predictive criterion for each study, approximated by $10$
independent Monte Carlo runs. The mean plus and minus two standard
deviations of these $10$ runs are shown in error bars. It is apparent
that the predictive power of the cut approach is higher for the first
study. For the other studies, the cut approach and the full posterior
give mostly comparable results. The graph helps identifying which study is
problematic for the full posterior approach. 

\section{Computational challenges\label{sec:Computational-challenges}}

In this section, we explain how the tables and figures of Section
\ref{sec:Numerical-experiments} were obtained, and the associated
computational challenges. We first discuss algorithms to sample 
distributions and to estimate their normalizing constants (Section \ref{subsec:Sampling-and-constants}).
We then discuss the challenges associated with the cut distribution (Section \ref{subsec:Sampling-from-cut}).
Finally, we revisit computational issues in the context of data confidentiality
(Section \ref{subsec:Computational-confidentiality}). 

\subsection{Sampling distributions and estimating their normalizing constants \label{subsec:Sampling-and-constants}}

We assume that the prior density and the likelihood
can be evaluated for each parameter, up to a multiplicative constant,
in both modules. Generic Monte Carlo algorithms can be used to obtain
draws from various candidate distributions, such as $\bar{\pi}(\theta_{1}|Y_{1})$
or $\bar{\pi}(\theta_{1},\theta_{2}|Y_{1},Y_{2})$. In order to compare
modular approaches, we require estimates of the associated predictive
scores, such as $\log\bar{\pi}(Y_{1})$ in the first module, or $\log\bar{\pi}(Y_{1}|Y_{2})$
in the full model. These scores correspond to logarithms of normalizing
constants in Bayes' formula, e.g. $\bar{\pi}(\theta_{1}|Y_{1})=p_{1}(\theta_{1})p(Y_{1}|\theta_{1})/\bar{\pi}(Y_{1})$,
$\bar{\pi}(\theta_{1},\theta_{2}|Y_{1},Y_{2})=\bar{\pi}(\theta_{1},\theta_{2}|Y_{2})p_{1}(Y_{1}|\theta_{1})/\bar{\pi}(Y_{1}|Y_{2})$.
Importance sampling and sequential Monte Carlo (SMC) samplers \citep{chopin2002sequential,DelDouJas:2006}
approximate posterior distributions and estimate their normalizing constants jointly.
Theoretical works provide support and insight on their precision as a function of their
computational cost \citep{schweizer2012non,whiteley2012sequential,beskos2014stability,beskos2014error},
and SMC samplers have been shown to compare favorably to other normalizing
constant estimation techniques in \citet{zhou2016toward}. Details
on the adaptive SMC sampler used in all experiments are given in Appendix \ref{app:smcdetails}.

\subsection{Sampling from the cut distribution \label{subsec:Sampling-from-cut}}

The approximation of the cut distribution $\pi^{\text{cut}}(\theta_{1},\theta_{2}|Y_{1},Y_{2})$
and its predictive score, as in Eq. \eqref{eq:prequential:partialfeedback},
can also be done with SMC samplers, in two stages. First, one obtains
a sample $\Theta_{1}^{1:N}$ approximating $\bar{\pi}(\theta_{1}|Y_{1})$.
To approximate the predictive score of the cut approach, one can approximate
each term in the sum of Eq. \eqref{eq:prequential:partialfeedback},
\[
\log(\int p_{2}(Y_{2}^{i}|\theta_{1},\theta_{2})\bar{\pi}(\theta_{1}|Y_{1})\bar{\pi}(\theta_{2}|\theta_{1},Y_{2}^{1:i-1})d\theta_{1}d\theta_{2}),
\]
for $1\leq i\leq n_{2}$, by the Monte Carlo approximation $\log(N^{-1}\sum_{k=1}^{N}\bar{\pi}(Y_{2}^{i}|\Theta_{1}^{k},Y_{2}^{1:i-1}))$.
The terms $\bar{\pi}(Y_{2}^{i}|\Theta_{1}^{k},Y_{2}^{1:i-1})$, for
$1\leq i\leq n_{2}$ can be approximated, for each $\Theta_{1}^{k}$,
$1\leq k\leq N$, again by Monte Carlo. This yields
a sample $\Theta_{2,i-1}^{k,1:M}$ approximating
$\bar{\pi}(\theta_{2}|\Theta_{1}^{k},Y_{2}^{1:i-1})$, for some integer $M$
chosen by the user. One can prune these samples to obtain one value $\Theta_{2,i-1}^k$,
for each $k$; the
resulting pairs $(\Theta_{1}^{k},\Theta_{2,i-1}^{k})_{k=1}^{N}$ approximate
the joint cut distribution given $Y_1$ and $Y_2^{1:i-1}$. Iterating through
the data $Y_2^1,\ldots,Y_2^{n_2}$ yields approximations of the sequence of cut distributions
and of the cut score of Eq. \eqref{eq:prequential:partialfeedback}.

The above procedure can be considered naive since it involves running a Monte Carlo 
method for each value $\Theta_{1}^{k}$ obtained in the approximation of $\bar{\pi}(\theta_{1}|Y_{1})$.
Although these runs can be done in parallel, it raises the question
of whether a sampling approach targeting the cut distribution is possible
in one stage. \citet{plummer2014cuts} discusses the problem of designing
Markov kernels leaving $\pi^{\text{cut}}(\theta_{1},\theta_{2}|Y_{1},Y_{2})$
invariant, and notes that previous attempts have resulted in incorrect
samplers. The main difficulty lies in the intractability of the term $\bar{\pi}(Y_{2}|\theta_{1})$ in Eq.
\eqref{eq:jointcut}, which involves an integral over $\theta_2$.
An alternative way of approximating the cut distribution
is proposed in \citet{jacob2017debiased}.

\subsection{Two-stage Monte Carlo approaches in confidential settings\label{subsec:Computational-confidentiality}}

A two-stage Monte Carlo approach, such as that described in the previous section, has
practical advantages in a context of confidential data sets. For various
reasons, the user might not have access simultaneously to $Y_{1}$
and $Y_{2}$, or even to both prior and likelihood functions. Consider
the scenario where one is given draws $\Theta_{1}^{1:N}$ from some
distribution $\bar{\pi}(\theta_{1}|Y_{1})$, which could be the posterior
in a first module, and one wants to consider modular and full approaches
for a second module. The cut distribution and its predictive score
can be approximated via the above SMC procedures. Can one retrieve
the full posterior distribution $\bar{\pi}(\theta_{1},\theta_{2}|Y_{1},Y_{2})$,
without further access to the first module?

As described above, algorithms such as SMC samplers provide $\Theta_{2,i}^{k}$ approximately from $\bar{\pi}(\theta_{2}|\Theta_{1}^{k},Y_{2}^{1:i})$,
as well as estimates $\hat{\pi}^{M}(Y_{2}^{i}|\Theta_{1}^{k},Y_{2}^{1:i-1})$
of the terms $\bar{\pi}(Y_{2}^{i}|\Theta_{1}^{k},Y_{2}^{1:i-1})$,
for all $1\leq i\leq n_{2}$. This enables an importance sampling
scheme, where the proposal distribution is $\pi^{\text{cut}}(\theta_{1},\theta_{2}|Y_{1},Y_{2}^{1:i})$
and the target distribution is $\bar{\pi}(\theta_{1},\theta_{2}|Y_{1},Y_{2}^{1:i})$,
for any $1\leq i\leq n_{2}$. The importance weight function is the
ratio of these densities, proportional $\bar{\pi}(Y_{2}^{1:i}|\theta_{1})$
for all $\theta_{1},\theta_{2}$, as can be seen from Eq. \eqref{eq:jointcut}. Therefore, we can define the importance
weights as $w_{i}^{k}=\hat{\pi}^{M}(Y_{2}^{1:i}|\Theta_{1}^{k})=\prod_{s=1}^{i}\hat{\pi}^{M}(Y_{2}^{s}|\Theta_{1}^{k},Y_{2}^{1:s-1})$,
and normalize the weights as $W_{i}^{k}=w_{i}^{k}/\sum_{j=1}^{N}w_{i}^{j}$.
The weighted samples $(W_{i}^{k},(\Theta_{1}^{k},\Theta_{2,i}^{k}))$,
for $1\leq k\leq N$, approximate the distribution
$\bar{\pi}(\theta_{1},\theta_{2}|Y_{1},Y_{2}^{1:i})$.

Furthermore, we can also approximate the predictive score $\log\bar{\pi}(Y_{2}|Y_{1})$,
using the decomposition
\begin{align*}
\log\bar{\pi}(Y_{2}|Y_{1}) & =\sum_{i=1}^{n_{2}}\log\bar{\pi}(Y_{2}^{i}|Y_{1},Y_{2}^{1:i-1})=\sum_{i=1}^{n_{2}}\log\left(\int p_{2}(Y_{2}^{i}|\theta_{1},\theta_{2})\bar{\pi}(\theta_{1},\theta_{2}|Y_{1},Y_{2}^{1:i-1})d\theta_{1}d\theta_{2}\right).
\end{align*}
The integral on the right hand side can be approximated by the weighted
samples introduced above. The idea of using 
Monte Carlo samples of a first distribution in an algorithm targeting
a full posterior distribution can be found in \citet{lunn:2013},
where the samples obtained in a first-stage MCMC chain are used as
proposals in a second stage. The method of \citet{lunn:2013}
could be applied here, but would not directly provide estimates
of the normalizing constants, which are required for our proposed
predictive scores.

The above approach is expected to work insofar as $\pi^{\text{cut}}(\theta_{1},\theta_{2}|Y_{1},Y_{2})$
is efficient enough as an importance sampling proposal for the target $\bar{\pi}(\theta_{1},\theta_{2}|Y_{1},Y_{2})$.
In other words, the importance sampler will
work if the feedback of $Y_{2}$ on $\theta_{1}$ is mild. Indeed
a mild feedback implies that the first stage samples $\Theta_{1}^{1:N}$
already are a good approximation of the marginal of $\theta_{1}$
in the full posterior. On the other hand, if the feedback
of $Y_{2}$ on $\theta_{1}$ is strong, approximating the marginal
$\bar{\pi}(\theta_{1}|Y_{1},Y_{2})$ with samples from $\bar{\pi}(\theta_{1}|Y_{1})$
will likely fail. 
Figure \ref{fig:biaseddata:marginalparameters} illustrates
the potential mismatch between these distributions. In case of such mismatch,
a sample $\Theta_{1}^{1:N}$ from the first module's posterior does
not spread enough over the support of the full posterior,
which means that the prior $p_{1}(\theta_{1})$ and the likelihood
$p_{1}(Y_{1}|\theta_{1})$ of the first module will have to be accessed
again. This motivates the search for computational
methods which would approximate the full posterior in two stages
while querying the first module as rarely as possible.

\section{Discussion \label{sec:Discussion}}

Combining task-specific data and models into coherent ensembles will be
instrumental in the understanding of uncertainty in large-scale systems,
arising in all fields: for instance,
medical models of the human body and its organs \citep{Dance02062015},
models of our planet and its climate systems \citep{Shen:climatesupermodels}, and
models of ecosystems and their interacting species \citep{Collie:ecosystem}. 
How misspecified components should interact, and whether data can be used
to derive optimal assemblies of components, will become pressing questions. In
various settings, scientists have resorted to modular approaches to deal with 
misspecification while propagating uncertainties. 
We propose a principled and data-driven procedure to help 
deciding between modular and full inferential approaches, 
to make the best use of the available modules. 
As illustrated in the meta-analysis example in Section \ref{sec:metaanalysis},
the proposed criteria can be modified on a case-by-case basis to accommodate
specificities, such as improper priors.

In numerical experiments, the proposed framework confirms that modular
approaches outperform the full model posterior distribution in various
settings, including meta-analysis and causal inference with propensity scores.
The proposed plan of action relies on predictive scores, and on the selection of
candidate distributions that yield the best predictions within the module that
gives an interpretable meaning to the parameters, as described in Section
\ref{sub:predictionplan}. 

Our proposed cut score depends on the ordering of the observations, which is
potentially problematic; averaging over permutations of the data alleviates the
issue but could be computationally expensive. 
Furthermore, the approximation of the cut distribution itself is challenging,
as mainstream MCMC algorithms cannot be used, due to the intractability of
the feedback term \citep{plummer2014cuts,jacob2017debiased}. Other modular approaches, such as those using power likelihoods
mentioned in Section \ref{sec:Why-not-fullposterior}, would allow
partial feedback of some modeling components on others; they would 
however raise their own computational challenges.

\section*{Acknowledgements}
The authors are thankful to Luke Bornn, Tristan Gray--Davies, Geoff Nicholls, Aimee Taylor and James Watson
for stimulating discussions.
Pierre E. Jacob gratefully acknowledges support by the National Science Fundation through grant DMS-1712872.
Chris C. Holmes gratefully acknowledges funding from the EPSRC and the MRC UK. Christian Robert is supported by
a 2016--2021 Institut Universitaire de France grant.

\bibliographystyle{abbrvnat}
\bibliography{biblio}

\appendix

\section{Calculations for the meta-analysis example \label{app:meta:calculations}}

To prove Eq. (\ref{eq:posteriorsigma_i}), we proceed as follows.
By noting that 
\[
\sum_{j=1}^{n_{i}}\left(y_{i,j}-b_{i}\right)^{2}=\sum_{j=1}^{n_{i}}\left(y_{i,j}-\bar{y}_{i}\right)^{2}+n_{i}\left(\bar{y}_{i}-b_{i}\right)^{2},
\]
we obtain 
\[
p(y_{i}|b_{i},\sigma_{i}^{2})=\left(2\pi\sigma_{i}^{2}\right)^{-n_{i}/2}\exp\left(-\frac{1}{2\sigma_{i}^{2}}\left(\sum_{j=1}^{n_{i}}\left(y_{i,j}-\bar{y}_{i}\right)^{2}+n_{i}\left(\bar{y}_{i}-b_{i}\right)^{2}\right)\right).
\]
This leads to the posterior of $b_{i},\sigma_{i}^{2}$ given $\tau^{2}$
and $y_{i}$:
\begin{equation}
p(b_{i},\sigma_{i}^{2}|\tau^{2},y_{i})\propto\left(\sigma_{i}^{-2}\right)\sigma_{i}^{-n_{i}}\exp\left(-\frac{1}{2\sigma_{i}^{2}}\left(\sum_{j=1}^{n_{i}}\left(y_{i,j}-\bar{y}_{i}\right)^{2}+n_{i}\left(\bar{y}_{i}-b_{i}\right)^{2}\right)-\frac{1}{2\tau^{2}}b_{i}^{2}\right).\label{eq:posteriorbisigmai}
\end{equation}
We can integrate $b_{i}$ out. Considering only the term in the exponential
that features $b_{i}$, we note
\begin{align*}
\frac{n_{i}(\bar{y}_{i}-b_{i})^{2}}{\sigma_{i}^{2}}+\frac{b_{i}^{2}}{\tau^{2}} & =\frac{n_{i}(\bar{y}_{i})^{2}-2b_{i}n_{i}\bar{y}_{i}+n_{i}b_{i}^{2}}{\sigma_{i}^{2}}+\frac{b_{i}^{2}}{\tau^{2}}\\
 & =(\frac{1}{\tau^{2}}+\frac{n_{i}}{\sigma_{i}^{2}})b_{i}^{2}-2\left(\frac{n_{i}\bar{y}_{i}}{\sigma_{i}^{2}}\right)b_{i}+\frac{n_{i}\left(\bar{y}_{i}\right)^{2}}{\sigma_{i}^{2}}\\
 & =(\frac{1}{\tau^{2}}+\frac{n_{i}}{\sigma_{i}^{2}})\left(b_{i}-\left(\frac{1}{\tau^{2}}+\frac{n_{i}}{\sigma_{i}^{2}}\right)^{-1}\left(\frac{n_{i}\bar{y}_{i}}{\sigma_{i}^{2}}\right)\right)^{2}-\left(\frac{1}{\tau^{2}}+\frac{n_{i}}{\sigma_{i}^{2}}\right)^{-1}\left(\frac{n_{i}\bar{y}_{i}}{\sigma_{i}^{2}}\right)^{2}+\frac{n_{i}\left(\bar{y}_{i}\right)^{2}}{\sigma_{i}^{2}}.
\end{align*}
Write $\left(\frac{1}{\tau^{2}}+\frac{n_{i}}{\sigma_{i}^{2}}\right)^{-1}=\frac{\tau^{2}\sigma_{i}^{2}}{n_{i}\tau^{2}+\sigma_{i}^{2}}$.
Thus, integrating out $b_{i}$ leads to 
\[
\int\exp\left(-\frac{1}{2}(\frac{1}{\tau^{2}}+\frac{n_{i}}{\sigma_{i}^{2}})\left(b_{i}-\left(\frac{1}{\tau^{2}}+\frac{n_{i}}{\sigma_{i}^{2}}\right)^{-1}\left(\frac{n_{i}\bar{y}_{i}}{\sigma_{i}^{2}}\right)\right)^{2}\right)db_{i}=\sqrt{2\pi\frac{\tau^{2}\sigma_{i}^{2}}{n_{i}\tau^{2}+\sigma_{i}^{2}}}.
\]
Furthermore,
\[
-\left(\frac{1}{\tau^{2}}+\frac{n_{i}}{\sigma_{i}^{2}}\right)^{-1}\left(\frac{n_{i}\bar{y}_{i}}{\sigma_{i}^{2}}\right)^{2}+\frac{n_{i}\left(\bar{y}_{i}\right)^{2}}{\sigma_{i}^{2}}=-\frac{\tau^{2}}{n_{i}\tau^{2}+\sigma_{i}^{2}}\frac{(n_{i}\bar{y}_{i})^{2}}{\sigma_{i}^{2}}+\frac{n_{i}\left(\bar{y}_{i}\right)^{2}}{\sigma_{i}^{2}}=\frac{n_{i}\tau^{2}+\sigma_{i}^{2}-n_{i}\tau^{2}}{\left(n_{i}\tau^{2}+\sigma_{i}^{2}\right)\sigma_{i}^{2}}n_{i}\left(\bar{y}_{i}\right)^{2}=\frac{\left(\bar{y}_{i}\right)^{2}}{\left(\tau^{2}+\sigma_{i}^{2}/n_{i}\right)},
\]
 and so 
\begin{align*}
p(\sigma_{i}^{2}|\tau^{2},y_{i}) & \propto\sigma_{i}^{-2}\sigma_{i}^{-n_{i}}\frac{\sigma_{i}}{\sqrt{\tau^{2}+\sigma_{i}^{2}/n_{i}}}\exp\left(-\frac{1}{2}\frac{\left(\bar{y}_{i}\right)^{2}}{\left(\tau^{2}+\sigma_{i}^{2}/n_{i}\right)}-\frac{1}{2\sigma_{i}^{2}}\left(\sum_{j=1}^{n_{i}}\left(y_{i,j}-\bar{y}_{i}\right)^{2}\right)\right)\\
 & \propto\frac{\sigma_{i}^{-n_{i}-1}}{\sqrt{\tau^{2}+\sigma_{i}^{2}/n_{i}}}\exp\left(-\frac{1}{2}\frac{\left(\bar{y}_{i}\right)^{2}}{\left(\tau^{2}+\sigma_{i}^{2}/n_{i}\right)}-\frac{s_{i}^{2}}{2\sigma_{i}^{2}}\right),
\end{align*}
with $s_{i}^{2}=\sum_{j=1}^{n_{i}}\left(y_{i,j}-\bar{y}_{i}\right)^{2}$.
This is Eq. (\ref{eq:posteriorsigma_i}). Eq. (\ref{eq:posteriortausigma})
follows by taking the product over the studies, and multiplying by the
prior on $\tau^{2}$ given $\sigma_{1:N}^{2}$.

To prove Eq. (\ref{eq:bsgivenrest}), we proceed as follows, starting
from Eq. (\ref{eq:posteriorbisigmai}). Completing the square leads
to the calculation
\begin{align*}
 & \exp\left(-\frac{1}{2\sigma_{i}^{2}}n_{i}\left(\bar{y}_{i}-b_{i}\right)^{2}-\frac{1}{2\tau^{2}}b_{i}^{2}\right)\\
\propto & \exp\left(-\frac{1}{2}(b_{i}^{2}(n_{i}\sigma_{i}^{-2}+\tau^{-2})-2n_{i}\sigma_{i}^{-2}b_{i}\bar{y}_{i})\right)\\
\propto & \exp\left(-\frac{1}{2}(n_{i}\sigma_{i}^{-2}+\tau^{-2})\left(b_{i}^{2}-2\frac{n_{i}\sigma_{i}^{-2}}{n_{i}\sigma_{i}^{-2}+\tau^{-2}}b_{i}\bar{y}_{i}\right)\right),
\end{align*}
where we recognize a Normal density with mean $n_{i}\sigma_{i}^{-2}/(n_{i}\sigma_{i}^{-2}+\tau^{-2})\bar{y}_{i}=\bar{y}_{i}\tau^{2}/(\tau^{2}+\sigma_{i}^{2}/n_{i})$,
and variance $1/(n_{i}\sigma_{i}^{-2}+\tau^{-2})=(\tau^{2}\sigma_{i}^{2}/n_{i})/(\tau^{2}+\sigma_{i}^{2}/n_{i})$.

Finally, we compute $p(y_{i,j+1}|\sigma_{i}^{2},\tau^{2},y_{i,1},\ldots,y_{i,j})$,
a quantity appearing in the proposed predictive criteria. Noting
that 
\[
p(y_{i,j+1}|\sigma_{i}^{2},\tau^{2},y_{i,1},\ldots,y_{i,j})=\int p(y_{i,j+1}|b_{i},\sigma_{i}^{2})p(b_{i}|\sigma_{i}^{2},\tau^{2},y_{i,1},\ldots,y_{i,j})db_{i},
\]
and using Eq. (\ref{eq:bsgivenrest}) and $y_{i,j+1}|b_{i},\sigma_{i}^{2}\sim\mathcal{N}(b_{i},\sigma_{i}^{2})$,
we have 
\[
y_{i,j+1}|\sigma_{i}^{2},\tau^{2},y_{i,1},\ldots,y_{i,j}\sim\mathcal{N}\left(\frac{j^{-1}\sum_{k=1}^{j}y_{i,k}\tau^{2}}{\tau^{2}+\sigma_{i}^{2}/j},\frac{\tau^{2}\sigma_{i}^{2}/j}{\tau^{2}+\sigma_{i}^{2}/j}+\sigma_{i}^{2}\right).
\]

\section{Adaptive Sequential Monte Carlo samplers\label{app:smcdetails}}

We first describe a generic adaptive SMC sampler, that starts from 
$N$ samples $\Theta_0^1,\ldots\Theta_0^N$ distributed according
to $p_0$, and produces weighted samples $(W_1^k,\Theta_1^k)_{k=1}^N$
approximately distributed according to $p_1$, along with an estimator
of the ratio of normalizing constant $Z_1/Z_0$, where
$Z_i = \int p^u_i(\theta)d\theta$, and $p^u_i$ denotes the unnormalized
probability density function associated with $p_i$. One can then reduce the samples
to one value by drawing an index $K$ with probabilities $W_1^{1:N}$ and returning $\Theta_1^K$.

We introduce, for any $\gamma \in [0,1]$, the distribution
$p_\gamma$ with unnormalized density function
$\theta \mapsto p_0(\theta)^{1-\gamma} p_1(\theta)^\gamma$. Its normalizing constant is denoted by $Z_\gamma$. Assuming that
we have a sample $(W_t^k,\Theta_t^k)_{k=1}^N$
approximately distributed according to $p_{\gamma_t}$, we proceed as follows to obtain $\gamma_{t+1}$ and $(W_{t+1}^k,\Theta_{t+1}^k)_{k=1}^N$,
approximately distributed according to $p_{\gamma_{t+1}}$. Note that
$p_{\gamma_{t+1}}(\theta)/p_{\gamma_{t}}(\theta) = (p_1(\theta)/p_0(\theta))^{\gamma_{t+1}-\gamma_t}$.

\begin{enumerate}
    \item For each $k\in 1:N$, compute $g_t^k = p_1(\Theta_t^k)/p_0(\Theta_t^k)$.
    \item Find the largest $\gamma_{t+1} \in [\gamma_t,1]$ such that some criterion depending on $(\gamma_t, \gamma_{t+1}, g_t^{1:N}, \Theta_t^{1:N})$ is met;
        see below for examples.
    \item For that $\gamma_{t+1}$, compute $w_{t+1}^k = (g_t^k)^{\gamma_{t+1}-\gamma_t}$ and $W_{t+1}^k = w_{t+1}^k / \sum_{j=1}^N w_{t+1}^j$.
    \item Compute $\sum_{k=1}^N W_t^k  w_{t+1}^k$, which approximates $Z_{\gamma_{t+1}}/Z_{\gamma_t}$.
    \item Either set $\Theta_{t+1}^k = \Theta_t^k$ for each $k\in 1:N$ and output $(W_{t+1}^k,\Theta_{t+1}^k)_{k=1}^N$,
        or perform a rejuvenation step: resample the particles according to the weights $W_{t+1}^{1:N}$, re-define $W_{t+1}^k = N^{-1}$, 
        and run, for each $\Theta_t^k$, a Markov kernel leaving $p_{\gamma_{t+1}}$ invariant to obtain each $\Theta_{t+1}^k$;
        see below for examples.
\end{enumerate}

Starting from $p_0$, we can perform the above procedure until $\gamma_{t+1} = 1$, in which case the algorithm terminates
with a sample approximately distributed according to $p_1$. The estimator of $Z_1/Z_0$ is obtained as a product of
all intermediate estimators of $Z_{\gamma_{t+1}}/Z_{\gamma_t}$. As an illustration, suppose that we want to approximate a
posterior distribution $p(\theta|Y^1,\ldots,Y^n)$ and its constant $p(Y^1,\ldots,Y^n)$, starting from draws from the prior distribution $p(\theta)$. 
We can decide to define $p_0$ as the prior and $p_1$ as the posterior. Alternatively, we can define $p_1$ to be the posterior 
given one observation, $p(\theta| Y^1)$. Then, we can run the procedure again with $p_0$ being $p(\theta| Y^1)$ and $p_1$ being $p(\theta| Y^1,Y^2)$,
and iterate until all observations have been assimilated. This is the strategy employed in our numerical experiments.

As a generic choice of Markov kernel for the rejuvenation step, we use
an independent Metropolis--Hastings kernel with a proposal distribution taken
to be a mixture of multivariate Normals. The mixture is calibrated on the current particles
before the rejuvenation step; we have used five components as a default choice.  

As a generic choice of adaptation rule, for the second step of the above
algorithm, we proceed as follows.  We define a diversity parameter $\alpha$,
set to $0.5$ by default, which indicates the desired minimum proportion of
unique particles within our sample, at all steps. The idea is to find
$\gamma_{t+1}$ such that, upon resampling the particles using weights
$W_{t+1}^k \propto (g_t^k)^{\gamma_{t+1}-\gamma_t}$, we obtain at least a
proportion $\alpha$ of unique particles.  This is complicated by the randomness
of the resampling step. Therefore, we clamp that randomness by drawing the
uniform variables of the resampling step and keeping them fixed during the
calculation of $\gamma_{t+1}$.  The proportion of unique particles after
resampling is then a deterministic function of $\gamma_{t+1}$, denoted by
$f(\gamma_{t+1})$.  We use a numerical optimizer to minimize $|f(\gamma_{t+1})
- \alpha|$ over $\gamma_{t+1}\in [\gamma_t,1]$.  For the resampling step, we
use systematic resampling, which involves only one uniform random variable at
each step. Theoretical support for adaptive SMC samplers has been studied in \citet{delmo:2012};
see also \citet{fearnhead2013adaptive,zhou2016toward}.

\end{document}

%% file: biaseddata.predictY1.tex
\begin{tabular}{ll}
  \hline
 & predictive score on $Y_1$ \\ 
  \hline
module 1 & -144.5 [-144.5, -144.4] \\ 
  prior & -151.4 [-152, -150.6] \\ 
  full model & -165 [-165.1, -165] \\ 
  module 2 & -188.8 [-189.1, -188.4] \\ 
   \hline
\end{tabular}

%% file: biaseddata.predictY2.tex
\begin{tabular}{ll}
  \hline
 & predictive score on $Y_2$ \\ 
  \hline
module 2 & -1402.7 [-1402.8, -1402.7] \\ 
  full model & -1423.2 [-1423.2, -1423.1] \\ 
  cut approach & -1443.4 [-1443.8, -1442.8] \\ 
  plug-in approach & -1443.4 [-1443.8, -1442.9] \\ 
   \hline
\end{tabular}

%% file: plummer.predictY1.tex
\begin{tabular}{ll}
  \hline
 & predictive score on $Y_1$ \\ 
  \hline
module 1 & -64.8 [-65, -64.7] \\ 
  full model & -74.9 [-75.1, -74.5] \\ 
  module 2 & -262.7 [-276.6, -253.5] \\ 
   \hline
\end{tabular}

%% file: plummer.predictY2.tex
\begin{tabular}{ll}
  \hline
 & predictive score on $Y_2$ \\ 
  \hline
module 2 & 11289 [11288.9, 11289.1] \\ 
  full model & 11285.1 [11285, 11285.2] \\ 
  cut approach & 11176 [11147.6, 11218.2] \\ 
  plug-in approach & 10836.2 [10836.1, 10836.3] \\ 
   \hline
\end{tabular}

%% file: propensity.predictY1.tex
\begin{tabular}{ll}
  \hline
 & predictive score on $X$ \\ 
  \hline
module 1 & -632.9 [-633.1, -632.7] \\ 
  full model & -664.4 [-665.5, -663.4] \\ 
  prior & -694.4 [-695.6, -693.1] \\ 
  module 2 & -6775.7 [-8422.4, -5351] \\ 
   \hline
\end{tabular}

%% file: propensity.predictY2.tex
\begin{tabular}{ll}
  \hline
 & predictive score on $Z$ \\ 
  \hline
module 2 & -624 [-624.5, -623.6] \\ 
  full model & -646.2 [-646.3, -646.1] \\ 
  cut approach & -648.7 [-650.3, -647] \\ 
  plug-in approach & -652.4 [-652.5, -652.3] \\ 
   \hline
\end{tabular}